\documentclass[fleqn,usenatbib]{mnras}

\usepackage{newtxtext,newtxmath}

\usepackage[T1]{fontenc}

\DeclareRobustCommand{\VAN}[3]{#2}
\let\VANthebibliography\thebibliography
\def\thebibliography{\DeclareRobustCommand{\VAN}[3]{##3}\VANthebibliography}

\usepackage{graphicx}	
\usepackage{amsmath}	
\usepackage{amssymb}	
\usepackage{caption}
\usepackage{hyperref}

\title[Modeling SSC and KN Effects in GRB Afterglows]{Modeling Synchrotron Self-Compton and Klein-Nishina Effects in Gamma-Ray Burst Afterglows}

\author[Jacovich, Beniamini, \& van der Horst]{
Taylor E. Jacovich,$^{1,2,3}$\thanks{E-mail: tjacovich@cfa.harvard.edu (TEJ)}
Paz Beniamini,$^{4,1,2}$
Alexander J. van der Horst$^{1,2}$
\\
$^{1}$Department of Physics, The George Washington University, 725 21st Street NW, Washington, DC 20052\\
$^{2}$Astronomy, Physics, and Statistics Institute of Sciences (APSIS), 725 21st Street NW, Washington, DC 20052\\
$^{3}$Smithsonian Astrophysical Observatory, 60 Garden Street, Cambridge, MA 02138\\
$^{4}$Division of Physics, Mathematics and Astronomy, California Institute of Technology, Pasadena, CA 91125}

\date{Accepted XXX. Received YYY; in original form ZZZ}

\pubyear{2020}

\begin{document}
\label{firstpage}
\pagerange{\pageref{firstpage}--\pageref{lastpage}}
\maketitle

\begin{abstract}
We present a self-consistent way of modeling synchrotron self-Compton (SSC) effects in gamma-ray burst afterglows, with and without approximated Klein-Nishina suppressed scattering. We provide an analytic approximation of our results, so that it can be incorporated into the afterglow modeling code \texttt{boxfit}, which is currently based on pure synchrotron emission.  We discuss the changes in spectral shape and evolution due to SSC effects, and comment on how these changes affect physical parameters derived from broadband modeling. We show that SSC effects can have a profound impact on the shape of the X-ray light curve using simulations including these effects. This leads to data that cannot be simultaneously fit well in both the X-ray and radio bands when considering synchrotron-only fits, and an inability to recover the correct physical parameters, with some fitted parameters deviating orders of magnitude from the simulated input parameters. This may have a significant impact on the physical parameter distributions based on previous broadband modeling efforts.
\end{abstract}




\section{Introduction}

Gamma-ray bursts (GRBs) are high energy bursts of $\gamma$-rays detected at cosmological distances isotropically across the sky. Since the first detection of a burst five decades ago, understanding of GRBs as the result of mechanisms internal to a relativistic jet driven by some central engine has been established \citep{RM92}. The jet interacts with the circumburst medium (CBM) producing the afterglow emission \citep{wrm97}. The canonical approach to afterglow modeling is to assume that a relativistic shock, formed at the jet-CBM interface, accelerates electrons into a power-law energy distribution; and those electrons radiate energy through the synchrotron emission process \citep{AG1,WG99}. This approach has proven remarkably successful at modeling afterglow emission, especially when coupled with hydrodynamic models of the jetted outflow \citep[e.g.][]{R99, CL99,PK02, AG2, vetajvdh}.

In spite of the successes of synchrotron-dominated afterglow models, many GRBs still remain resistant to characterization by this method. Evidence increasingly points towards additional emission mechanisms modifying, and at certain wave bands even dominating, the afterglow emission. In particular, the up-scattering of the original synchrotron emission off of the emitting electrons in synchrotron self-Compton (SSC) emission  occurs at some level in all afterglows; and it can become a dominant mechanism in both the emission and electron cooling processes, depending on the microphysical parameters related to the electrons and magnetic fields. SSC effects have been discussed in the literature, with great care taken to discuss all physical implications with the same level of detail as synchrotron emission in GRBs \citep[e.g.][]{se01,Nakar,Nava,beniamini}. In spite of this, SSC effects have been applied inconsistently to afterglow modeling, and mainly when GRB afterglows proved resistant to modeling with only synchrotron emission \citep[e.g.][]{CH1}. Many of these attempts adopted asymptotic descriptions of the emission and simplified the dynamics to that of a thin, symmetric, shell of homogeneous, relativistic material. These studies also tended to neglect the frequency-dependent suppression of SSC up-scattering through Klein-Nishina (KN) effects.

To improve the modeling of GRB afterglows, one needs a rigorous and generic way to describe the effect of SSC processes on the GRB spectrum based on the microphysics of the outflow. Doing this, one can consistently model all afterglows while taking into account both synchrotron and SSC effects. To accomplish this, we adopted the methodology of describing SSC effects with the SSC-to-synchrotron power ratio $Y$, which is a function of frequency and time. We derive equations for $Y$ in different spectral regimes and explain how it affects electron cooling. We introduce equations for $Y$ both without (\S \ref{SSCT}) and with (\S \ref{KNsec}) taking KN effects into account. We note that the inclusion of SSC effects can constrain afterglow physics without introducing any additional parameters compared to modeling with only synchrotron emission. The KN-approximated solution is coupled to the two-dimensional hydrodynamic afterglow modeling code \texttt{boxfit} \citep{vetajvdh} in \S \ref{KNimp}, so we can model light curves and spectra with an accurate consideration of the physics. With this new tool, we discuss the observed impact of SSC cooling on the spectral and temporal evolution of the afterglow with and without KN effects in \S \ref{effs}. We also discuss under what physical conditions these effects become relevant. Finally, we close with a discussion of what this means for past GRB modeling efforts in \S \ref{mod}, as well as future applications of this new addition to the afterglow modeling toolkit in \S \ref{sum}.

\section{SSC in The Thomson Regime}\label{SSCT}

\begin{table*}
\caption{$Y$ in the Thomson Regime}
\begin{tabular}{lcl}
\hline
Regime&Requirements&$Y_{c}$\\
\hline
\label{tab:Yt}
Fast-Cooling &$\gamma_{c} < \gamma_{m}$ &$ Y_{\rm T}(1+Y_{\rm T}) = \frac{(p-2)\epsilon_{\rm e}}{(p-1)\epsilon_{\rm B}}\left(\frac{p-1}{p-2}(1+Y_{\rm T})-\frac{\gamma_{\rm c}^s}{\gamma_{\rm m}}\right)\left((1+Y_{\rm T})-\frac{p-1}{p}\frac{\gamma_{\rm c}^s}{\gamma_{\rm m}}\right)^{-1}$ \\
Transition Value & $\gamma_{c}=\gamma_{m} $&$ Y_{\rm T}\equiv Y_* =\frac{1}{2}\left(\sqrt{1+\frac{4p}{p-1}\frac{\epsilon_{\rm e}}{\epsilon_{\rm B}}}-1\right)$ \\
Slow-Cooling & $\gamma_{c} > \gamma_{m}$ & $Y_{\rm T}(1+Y_{\rm T}) = \frac{p\left(\frac{\epsilon_{\rm e}}{\epsilon_{\rm B}}\frac{\gamma_{\rm m}}{\gamma_{\rm c}^s}(1+Y_{\rm T})^{p-1}\frac{p-2}{p-3}+\frac{\epsilon_{\rm e}}{\epsilon_{\rm B}}\frac{1}{3-p}\left(\frac{\gamma_{\rm m}}{\gamma_{\rm c}^s}\right)^{p-2}\right)}{ p(1+Y_{\rm T})^{p-1}-\left(\frac{\gamma_{\rm m}}{\gamma_{\rm c}^s}\right)^{p-1}}$\\
\hline
\end{tabular}
\end{table*}

The standard method for modeling emission from a GRB afterglow is to assume that electrons are instantaneously accelerated by a relativistic shock to a power-law distribution of Lorentz factor $\gamma_{\rm e}$, with a power-law slope $p$ above some minimum Lorentz factor $\gamma_{\rm m}$, which is a function of several microphysical parameters. These electrons are then assumed to radiate away energy through the synchrotron process with a characteristic frequency $\nu(\gamma_{\rm e}) = C_\nu \epsilon_{\rm B}^{1/2}\gamma_{\rm e}^2$ \citep{AG1, WG99}. The characteristic frequency depends on the total blastwave energy fraction in the magnetic field $\epsilon_{\rm B}$, as well as an overall constant, $C_\nu$, which has different values in the literature depending on how the above characteristic frequency was derived. In order to maintain consistency with \texttt{boxfit}, we chose $C_\nu$ and all other constants to match \citet{ve11}. A break in the distribution occurs at the Lorentz factor where electrons are radiating away significant portions of their energy on the dynamic timescale of the jet, denoted as the cooling Lorentz factor, $\gamma_{\rm c}$. We can write down the full electron distribution in the fast cooling ($\gamma_{\rm m} > \gamma_{\rm c}$) case:
\begin{equation}
\frac{dn_e^{\prime}}{d\gamma_{\rm e}} \propto \left\{
        \begin{array}{ll}
            \gamma_{\rm e}^{-2}, & \quad \gamma_{\rm c} < \gamma_{\rm e} \leq \gamma_{\rm m} \\
            \gamma_{\rm e}^
            {-p-1}, & \quad \gamma_{\rm e} > \gamma_{\rm m}
        \end{array}
    \right.
\end{equation}
or slow cooling ($\gamma_{\rm m} < \gamma_{\rm c}$) case:
\begin{equation}
\frac{dn_e^{\prime}}{d\gamma_{\rm e}} \propto \left\{
        \begin{array}{ll}
            \gamma_{\rm e}^{-p}, & \quad \gamma_{\rm m} < \gamma_{\rm e} \leq \gamma_{\rm c} \\
            \gamma_{\rm e}^
            {-p-1}, & \quad \gamma_{\rm e} > \gamma_{\rm c}
        \end{array}
    \right.
\end{equation}
The full set of parameters required to model the afterglow include those mentioned above ($\epsilon_{\rm B}$, $p$) as well as the fraction of the shock energy in the electron population, $\epsilon_{\rm e}$; the isotropic-equivalent energy of the jet, $E_{\text{iso}}$; the opening angle of the jet, $\theta_0$; the observer angle relative to the jet axis, $\theta_{\rm obs}$; the circumburst density, $n$ ($A$ in the case of a wind-like medium); and the fraction of accelerated electrons $\xi_N$ \citep{AG1, WG99, CL99}.

In many cases, this description of an afterglow is a reasonable approximation, but in cases where a small fraction of the shock energy is diverted into the magnetic field strength, i.e. $\epsilon_{\rm B}$ is small, synchrotron photon up-scattering begins to dominate electron cooling. This particular form of inverse-Compton emission is known as SSC emission. SSC effects present themselves in three distinct ways: the first is as increased electron cooling which leads to a lower value for $\gamma_{\rm c}$; the second is an overall decrease in synchrotron flux above $\nu_c$; and the third is increased emission for frequencies at and above $\sim \rm{min}(\gamma_{m},\gamma_{c})^2 \nu$ where $\nu$ is the seed photon frequency \citep{se01}. In practice, the cooling effects occur in the canonical observed afterglow regime, in particular in the X-ray band, while the SSC emission peak occurs at significantly higher energies. The effects of SSC can be incorporated into the synchrotron spectrum by solving the electron cooling equation assuming both synchrotron and SSC cooling, the result of which is 
\begin{equation}
\gamma_{\rm c} = \gamma_{c}^S\left(1+Y\right)^{-1}
\end{equation}
Here $\gamma_{c}^{S}$ is the cooling Lorentz factor assuming only synchrotron emission, while $\gamma_{c}$ is the effective cooling Lorentz factor. We have also introduced the SSC parameter $Y$ which relates the incident synchrotron power to the SSC emission. In the context of electron cooling, $Y$ is of importance to us only in cases where the total SSC power exceeds that of the synchrotron emission. We also note that $Y$ now appears explicitly in the cooling equation, which causes $\nu_{\rm c}$ to be reduced by  $(1+Y)^{-2}$. In the Thomson scattering case, $Y(\gamma_{\rm e}, t)$ reduces to $Y(t)$, which we denote as $Y_{\rm T}$ to differentiate it from the general $Y=Y(\gamma_{\rm e}, t)$.

To discuss $Y_{\rm T}$ in earnest, we must have a mathematical description of it based on the physical afterglow system.  It can be shown that in the Thomson regime, $Y_{\rm T}$ can be defined as \citep{beniamini}
\begin{equation}\label{Y}
Y_{\rm T} = \frac{4}{3}\sigma_{\rm T} n_0^{\prime} \Delta R^{\prime} \langle \gamma_{e}^2\rangle
\end{equation}

Here $\sigma_{\rm T}$ is the Thomson scattering cross-section, $n_0^{\prime}$ is the electron number density, $\Delta R^{\prime}$ is the length of a thin emitting shell at the shock boundary, and $\langle \gamma_{e}^2\rangle$ is the second moment of the electron Lorentz factor distribution. Primed values are calculated in the co-moving frame of the jet. $Y$ is frame invariant, so it is sufficient to determine it in only one frame. Although $Y$ has an apparent dependence on the size of the emitting region, $\Delta R^{\prime}$, the dependence ultimately cancels, and $Y$ becomes a function of the microphysical parameters (see appendices of this paper).  We present a full derivation of $Y$ in the Thomson regime in Appendix~\ref{appT}, and here we give a brief discussion of the three key regimes for the Thomson $Y$: fast cooling $(\gamma_{\rm c}<\gamma_{\rm m})$, slow cooling $(\gamma_{\rm c}>\gamma_{\rm m})$, and the transition where $\gamma_{\rm m} = \gamma_{\rm c}$. We present the resulting equations for $Y_{\rm T}$ in Table~\ref{tab:Yt}.

\subsection{Fast Cooling}
For $Y_{\rm T}$ in the fast-cooling regime, a closed-form solution can be constructed by solving a third order polynomial in $Y$. In the ultra-fast cooling approximation ($\gamma_{\rm c} \ll \gamma_{\rm m}$), we obtain 
\begin{equation}
Y_{\text{fast}}\propto \left\{
        \begin{array}{ll}
            \sqrt{\epsilon_{\rm e}\epsilon_{\rm B}^{-1}}, & \quad \epsilon_{\rm e}\epsilon_{\rm B}^{-1} \gg 1  \\
            \epsilon_{\rm e}\epsilon_{\rm B}^{-1}, & \quad \epsilon_{\rm e}\epsilon_{\rm B}^{-1} \ll 1
        \end{array}
    \right.
\end{equation}
As the function approaches the transition between regimes, terms proportional to $\gamma_{\rm c}\gamma_{\rm m}^{-1}$ become important, and the function rises steeply to the value calculated for the transition point. At the transition between fast and slow cooling, the two critical frequencies are identical, as are the respective electron distributions. The resulting equation for $Y$ reduces immensely, dropping all $\gamma_{\rm e}$ dependence, and giving a very clear depiction of the microphysical parameters that modify $Y_{\rm T}$. We denote this solution as $Y_{*}$.

\subsection{Slow Cooling}
\indent The slow-cooling regime requires more consideration than the previously discussed regimes. The electron distribution shares the same $\gamma_{\rm e}^{-p-1}$ behavior above the cooling break, but has a $\gamma_{\rm e}^{-p}$ dependence at lower energies. This dependence makes solving for $Y_{\rm T}$ in the slow cooling regime more difficult, as their is no closed form solution without knowing a priori what the power-law slope is. Even then, not all values of $p$ yield a closed solution. One exists for $p=2.5$, which we will make use of below, but generally $Y_{\rm T}$ must be solved numerically, and the computational costs of solving this equation in real-time are far higher than can be reasonably incorporated into \texttt{boxfit}, given that the code is meant to be used for iterative fitting of data. To work around this hurdle, we introduce an asymptotic solution to $Y_{\rm T}$ in the slow-cooling regime such that it returns approximately the right value at the transition between the cooling regimes, while also approximating the behavior in the limit $\gamma_{\rm c}\gg\gamma_{\rm m}$ well. The asymptotes are presented in Table~\ref{tab:Yslow}.

\begin{table*}
\caption{Approximating $Y_{\rm T}$ in the Slow-Cooling Regime}
\label{tab:Yslow}
\begin{tabular}{lcl}
\hline
Rule 1 & Rule 2 & $Y_{\rm slow}$\\
\hline
$\gamma_{m} \ll \gamma_{c}$ & & $Y_{\rm slow}(1+Y_{\rm slow})^{3-p}  \approx \frac{\epsilon_{\rm e}}{\epsilon_{\rm B}}\frac{1}{3-p}\left(\frac{\gamma_{\rm m}}{\gamma_{\rm c}^s}\right)^{p-2}$ \\

$\gamma_{m} \ll \gamma_{c}$ & $Y_{\rm slow} \gg 1$ &$Y_{\rm slow} \approx \left(\frac{\epsilon_{\rm e}}{\epsilon_{\rm B}}\frac{1}{3-p}\left(\frac{\gamma_{\rm m}}{\gamma_{\rm c}^s}\right)^{p-2}\right)^{\frac{1}{4-p}}$ \\

$\gamma_{m} \ll \gamma_{c}$ & $Y_{\rm slow} \ll 1$ & $Y_{\rm slow} \approx \left(\frac{\epsilon_{\rm e}}{\epsilon_{\rm B}}\frac{1}{3-p}\left(\frac{\gamma_{\rm m}}{\gamma_{\rm c}^s}\right)^{p-2}\right)$\\
\hline
\end{tabular}
\end{table*}


\section{Approximating KN Suppression of SSC Scattering}\label{KNsec}

At photon energies comparable to or larger than the electron rest mass in the electron center of mass frame, we can no longer assume a purely Thomson scattering cross-section. At these energies, electron recoil and KN suppression must be included to properly characterize the various emission mechanisms. KN suppression is particularly important as it has the effect of significantly reducing the electron scattering cross-section for high-energy synchrotron photons.  The exact behavior requires examining interactions at the individual particle level, but we can make two assumptions that greatly simplify the derived spectra, as was done in \citet{Nakar}. First we can assume that a given photon with frequency $\nu$ is in the Thomson regime for all electrons with Lorentz factor $\gamma$ for which $h \nu < \gamma_{\rm e} m_{\rm e}c^2$. Beyond this point, we can consider energy transfer as inconsequential, i.e. photons only gain a finite amount of energy proportional to the electron mass, as opposed to a squared Lorentz boost. We can denote the Lorentz factor of the maximum scattering electron, using the definition from \citet{Nakar}, as  
\begin{equation}
\hat{\gamma_{\rm e}}=\frac{m_{\rm e} c^2}{h \nu_{\rm e}}=\frac{m_{\rm e} c^2}{h C_{\nu}\epsilon_{\rm B}^{1/2}\gamma_{\rm e}^2} \propto \frac{1}{\gamma_{\rm e}^2}
\end{equation} 

We can then assume the scattering cross-section takes on a step function behavior such that
\begin{equation}
\sigma(\nu) = \left\{
        \begin{array}{ll}
            \sigma_{\rm T}, & \quad \gamma_{\rm e} \leq \hat{\gamma}\left(\nu\right) \\
            0, & \quad \gamma_{\rm e} > \hat{\gamma}\left(\nu\right)
        \end{array}
    \right.
\end{equation}
The modified scattering cross-section allows us to define a $\gamma_{\rm e}$-dependent description of $Y$, such that $Y$ transitions smoothly from $Y_{\rm T}$ described above to regimes where most electrons are beyond the KN limit for the observed photon frequency. A full description of KN effects on both the synchrotron and SSC emission spectra can be found in \citet{Nakar}. We will briefly discuss the pieces relevant for implementation in numerical codes such as \texttt{boxfit}.

The modified $Y$ parameter is derived from 
\begin{equation}\label{Ykn:eq}
Y\left(\gamma_{\rm e}\right) \ \propto \  \int_{0}^{\Tilde{\nu}\left(\gamma_{\rm e}\right)}d\nu'\int d\gamma_{\rm e}^* P_{\nu'}\left(\gamma_{\rm e}^*\right)\frac{dn_e^{\prime}}{d\gamma_{\rm e}^*}
\end{equation}
Unlike the Thomson case, where $Y$ can be shown to simplify to an integral over the electron population (see Appendix~\ref{appT}), $Y$ in the KN regime is defined as a convolution of the electron and photon populations, with a high-energy limit on the photon integral created by the step-function behavior of the scattering cross-section. We omit the overall constants associated with Equation~\ref{Ykn:eq}, because the results will ultimately be re-scaled such that they are consistent with the Thomson regime. We introduce the maximum frequency photon an electron can up-scatter in the Thomson regime $\Tilde{\nu}$, where
\[
\Tilde{\nu}=\nu_{\rm sync}\left(\tilde{\gamma}_{\rm e}\right)=C_\nu \epsilon_{\rm B}^{1/2} \tilde{\gamma}_{\rm e}^2
\]
Here, $\tilde{\gamma}_{\rm e}$ is the Lorentz factor of the electron that emitted the maximally scattered photon
\[
\tilde{\gamma}_{\rm e}=\left(\frac{\gamma_{\rm e} m_{\rm e} c^2}{h \nu_{\rm sync}\left(\gamma_{\rm e}\right)}
\right)^{1/2}=\left(\gamma_{\rm e}\hat{\gamma}_{\rm e}\right)^{1/2}\]
and 
\begin{equation}
    \Tilde{\nu}=C_\nu \epsilon_{\rm B}^{1/2}\left(\gamma_{\rm e}\hat{\gamma}_{\rm e}\right)
\end{equation}
Unlike the case in which we omit KN effects, we do not look for exact solutions for the various regimes as there is no simple way to merge them into a single equation that works for the entire GRB parameter space.  Instead, we determine the functional behavior of $Y$ as a function of $\gamma_{\rm e}$ in the KN-suppressed limit, and then self-consistently connect it to our Thomson-derived solution such that $Y$ is continuous in both time and frequency space.

\begin{table*}
\caption{Approximating $Y(\nu_{\rm e})$ for KN-Suppressed SSC}
\label{tab:YKN}
\begin{tabular}{lccl}
\hline
Regime&Rule 1 & Rule 2 & $Y(\nu_{\rm e})$\\
\hline
Fast-Cooling &$ \gamma_{c} < \gamma_{m} $&$ \gamma_{\rm e} < \hat{\gamma}_{m} $&$ Y(\nu_{\rm e}) = Y_{\rm T}$\\
 &$ \gamma_{c} < \gamma_{m} $&$\hat{\gamma}_{m} < \gamma_{\rm e} < \hat{\gamma}_{c}$ & $Y(\nu_{\rm e}) = Y_{\rm T} \left(\frac{\gamma_{e}}{\hat{\gamma}_{m}}\right)^{-\frac{1}{2}}$\\
  & $\gamma_{c} < \gamma_{m}$ & $\hat{\gamma}_{c} < \gamma_{e}$ &$ Y (\nu_{\rm e}) = Y_{\rm T}\frac{\gamma_{c}}{\gamma_{m}}\left(\frac{\gamma_{e}}{\hat{\gamma}_{c}}\right)^{-\frac{4}{3}}$ \\
  Slow-Cooling & $\gamma_{m} < \gamma_{c}$ & $\gamma_{\rm e} < \hat{\gamma}_{c}$ & $Y(\nu_{\rm e}) = Y_{\rm T}$ \\
  & $\gamma_{m} < \gamma_{c}$ &$\hat{\gamma}_{m} < \gamma_{\rm e} < \hat{\gamma}_{c}$ & $Y(\nu_{\rm e})=Y_{\rm T}\left(\frac{\gamma_{\rm e}}{\hat{\gamma}_{\rm c}}\right)^{\frac{p-3}{2}}$ \\
   & $\gamma_{m} < \gamma_{c}$ &$\gamma_{\rm e} < \hat{\gamma}_{m}$ &  $Y(\nu_{\rm e}) = Y_{\rm T}\left(\frac{\hat{\gamma}_{m}}{\hat{\gamma}_{\rm c}}\right)^{\frac{p-3}{2}}\left(\frac{\gamma_{e}}{\hat{\gamma}_{m}}\right)^{-\frac{4}{3}}$\\
   \hline
\end{tabular}
\end{table*}

\subsection{Fast Cooling}

In the fast-cooling case ($\gamma_{\rm m}>\gamma_{\rm c}$ or equivalently $\hat{\gamma}_{\rm c}>\hat{\gamma}_{\rm m}$), there are three key regimes worth discussing. The first is the Thomson regime discussed above; the next regime occurs when $\nu_{\rm{obs}}$ photons can no longer scatter off of $\gamma_{\rm m}$ electrons; and the final one occurs when those same photons can no longer scatter off of $\gamma_{\rm c}$ electrons. In practice, this results in a $Y$ parameter that goes as 
\begin{equation}
Y_{\text{fast}}\left(\gamma_{\rm e}\right)\  \propto\ \begin{cases} 
      \gamma_{\rm e}^0, & \gamma_{\rm e} \leq \hat{\gamma}_{\rm m} \\
     \gamma_{\rm e}^{-1/2}, & \hat{\gamma}_{\rm m} \leq \gamma_{\rm e} \leq \hat{\gamma}_{\rm c} \\
      \gamma_{\rm e}^{-4/3}, & \hat{\gamma}_{\rm c}  \leq \gamma_{\rm e}
   \end{cases}
\end{equation}
The full derivation of these regimes has already been carried out by \citet{Nakar}, so we will not repeat that here. We note that we omit several secondary regimes \citet{Nakar} defined as the power-law segments where $Y$ would be smaller than 1, as these cases would look identical to the synchrotron curve that we calculate independently. We do, however, include $Y \ll1$ asymptotes of the regime above to help with the transition from SSC- to synchrotron-dominated cooling.

\subsection{Slow Cooling}

The slow-cooling case ($\gamma_{\rm c}>\gamma_{\rm m}$ or equivalently $\hat{\gamma}_{\rm m}>\hat{\gamma}_{\rm c}$), presents itself in a more complicated fashion. Unlike fast cooling, the weakly-suppressed regime has a $p$ dependence. To deal with this, we again look to \citet{Nakar} as a basis for defining $Y$ in the KN regime, and use our Thomson solution to create a smoothed and continuous approximation for all times and frequencies:
\begin{equation}
Y_{\text{slow}}\left(\gamma_{\rm e}\right)\  \propto\ \begin{cases} 
      \gamma_{\rm e}^0, & \gamma_{\rm e} \leq \hat{\gamma}_{\rm c} \\
     \gamma_{\rm e}^{-(p-3)/2}, & \hat{\gamma}_{\rm c} \leq \gamma_{\rm e} \leq \hat{\gamma}_{\rm m} \\
      \gamma_{\rm e}^{-4/3}, & \hat{\gamma}_{\rm m}  \leq \gamma_{\rm e}
   \end{cases}
\end{equation}
In practice, $Y_{\rm slow}$ is further complicated by the strong $p$ dependence of $Y_{\rm T}$ in this regime, but we move any explicit discussion of that to Appendix~\ref{appT}. 
\subsection{Transition Between Cooling Regimes}
The low-energy regimes of both fast and slow cooling represent scattering of photons that are not the characteristic frequency of any electron, but are instead produced by the $\nu^\frac{1}{3}$ tail of the single electron spectrum. These photons are washed out at higher energies, but do appear at frequencies below $\nu($min$(\gamma_{\rm m},\gamma_{\rm c}))$. This places an important check on self-consistency, as the transitional regime removes the central power-law segment, and $Y_{\text{fast}}=Y_{\text{slow}}$ becomes
\begin{equation}
Y_{*}\left(\gamma_{\rm e}\right)  \propto\ \begin{cases} 
      \gamma_{\rm e}^0, & \gamma_{\rm e} \leq \hat{\gamma}_{\rm c} \\
      \gamma_{\rm e}^{-4/3}, & \hat{\gamma}_*  \leq \gamma_{\rm e}
   \end{cases}
\end{equation}

\section{Implementing SSC Effects into \lowercase{{\texttt{boxfit}}}} 
\label{KNimp}

\subsection{$Y$ in the Thomson Regime}

\begin{figure}
    \includegraphics[width=0.47\textwidth]{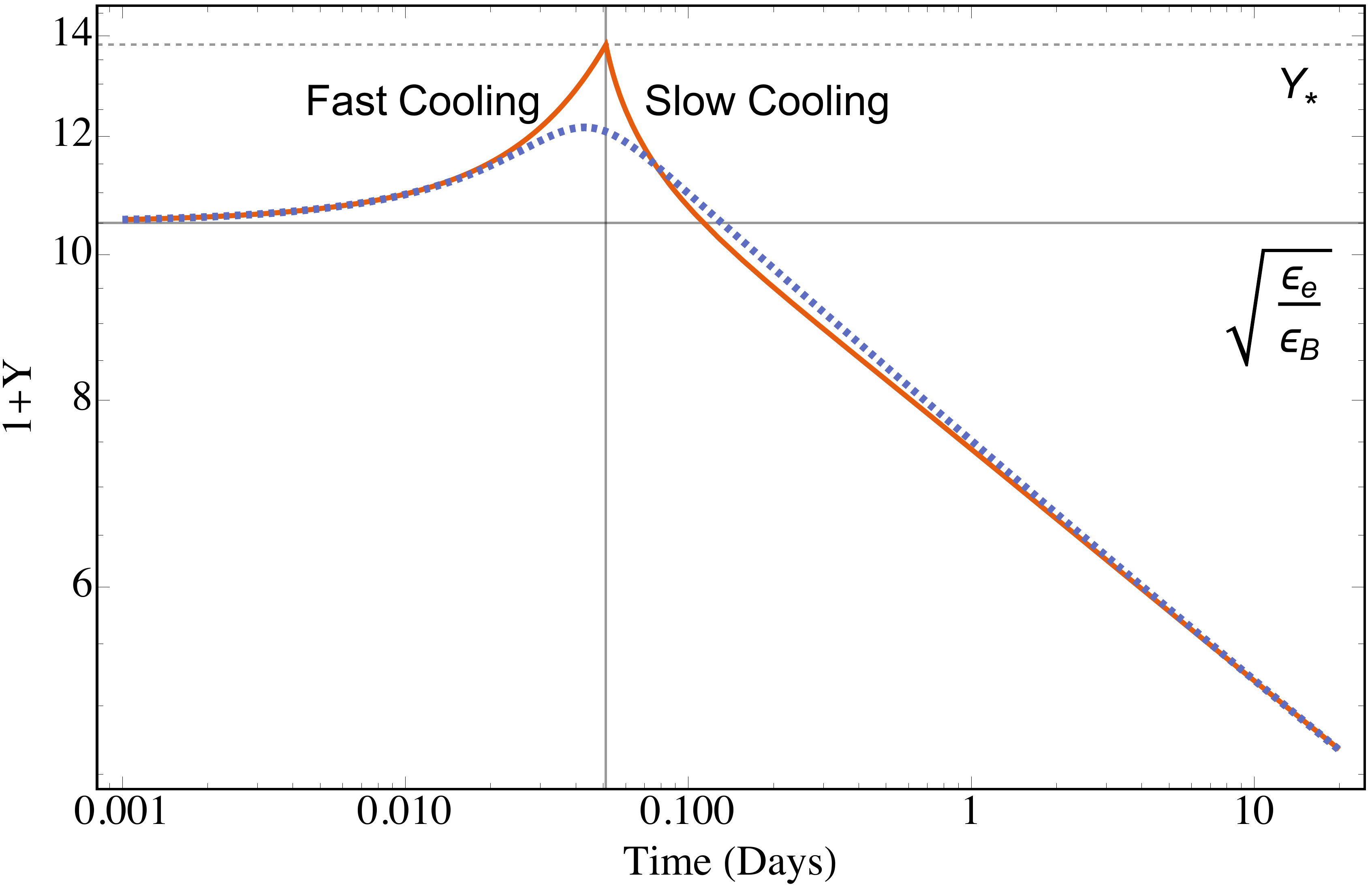}
    \centering\caption{ $Y$ as a function of time in the Thomson regime. The solid orange curve corresponds to the full solution for both cooling regimes with physical parameters $E_{\text{iso}}=10^{53}$ ergs, $n_0=5$ cm${}^{-3}$, $p=2.3$, $\epsilon_{\rm e}=10^{-1}$, $\epsilon_{\rm B}=10^{-3}$, $\theta_0=0.3$ rad, $\theta_{\rm obs}=0.0$.  The dashed, blue curve represents the exact $Y_{\text{fast}}$ solution smoothly connected to the $Y_{\text{slow}}$ approximation. The solid horizontal line denotes $\sqrt{\epsilon_{\rm e}\epsilon_{\rm B}^{-1}}$, and the dashed line corresponds to the transition value $Y_{*}$. The vertical line denotes the time for which $\gamma_{\rm c}= \gamma_{\rm m}$. }
    \label{Yfull}
\end{figure}

\indent Now that we have established the exact solution for $Y$ in each cooling regime, or at least how to obtain it, we can produce a solution for implementation in \texttt{boxfit}. There are two main issues that needed to be addressed, the first of which is the issue of the cooling regime. \texttt{boxfit} emission depends on the ordering of the critical frequencies and that ordering depends on $Y$. The problem arises because $Y$ also depends on the ordering and value of the critical frequencies. To alleviate this issue, we construct a smoothly broken power-law description of the form 
\begin{equation}
Y_{\rm T} = \left(Y_{\rm fast}^{\alpha}+Y_{\rm slow}^{\alpha}\right)^{1/\alpha}
\end{equation}
where $\alpha<0$.  Here, we are taking advantage of the fact that the two cooling regimes only intersect at the transition value, and that the two functions blow up rapidly outside of their own cooling regime, so we can select $\alpha$ such that $Y$ always selects the smaller of the two regimes. We find that a good fit for a large range of parameters is $\alpha=-60p^{-2}$; a visualization of this for $p=2.3$ is presented in Figure~\ref{Yfull}.  

Even with this approximation, we still need to deal with the numerical complexity of computing $Y_{\rm slow}$ for an afterglow with arbitrary parameters. We compute the asymptotic behavior of the slow-cooling case at late times, as was done in \citet{beniamini}. The result is a much simplified equation, but still not exactly solvable without cumbersome numerical techniques. To overcome this, we perform an additional smoothing of $Y_{\rm slow}$ using the limits for $Y\gg1$ and $Y\ll1$. The reduced solution and its asymptotes are also included in Appendix~\ref{appT}. The doubly-smoothed broken power law is plotted together with the exact solution in Figure~\ref{Yfull}. 

\begin{table*}
\centering
\caption{Approximating $Y(\nu_{c})$ for KN-Suppressed SSC}
\label{tab:Yckn}
\begin{tabular}{lcccl}
\hline
Regime&Rule 1&Rule 2&Rule 3&$Y_c$\\
\hline
Fast-Cooling &$ \gamma_{c} < \gamma_{m} $& $ \gamma_{\rm c} < \hat{\gamma}_{m} $& \ & $ Y(\nu_{c}) = Y_{\rm T}$\\
&$ \gamma_{c} < \gamma_{m} $ & $\hat{\gamma}_{m} < \gamma_{\rm c} < \hat{\gamma}_{c} $ & $ Y \gg 1$ & $ Y(\nu_{c}) = Y_{\rm T}^2 \left(\frac{\gamma_{c}^s}{\hat{\gamma}_{m}}\right)^{-1}$\\
&$ \gamma_{c} < \gamma_{m} $&$\hat{\gamma}_{m} < \gamma_{\rm c} < \hat{\gamma}_{c} $&$ Y \ll 1$&$ Y(\nu_{c}) = Y_{\rm T} \left(\frac{\gamma_{c}^s}{\hat{\gamma}_{m}}\right)^{-\frac{1}{2}}$\\
&$ \gamma_{c} < \gamma_{m}$ &$ \hat{\gamma}_{c} < \gamma_{c}$ & \ &$Y (\nu_{c}) = Y_{\rm T}\left(\gamma_{c}^s\right)^{-1}\hat{\gamma}_{m}^{\frac{1}{2}}$ \\
Slow-Cooling &$ \gamma_{m} < \gamma_{c}$ &$ \gamma_{\rm c} < \hat{\gamma}_{c}$& \ &$ Y(\nu_{c}) = Y_{\rm T} $\\
&$ \gamma_{m} \ll \gamma_{c}$ &$\hat{\gamma}_{m} < \gamma_{\rm c} < \hat{\gamma}_{c}$ &$Y\gg1$ &$Y(\nu_{c})=\left(\frac{\epsilon_{\rm e}}{\epsilon_{\rm B}(3-p)}\left(\frac{\gamma_{\rm m}}{\gamma_{\rm c}^s}\right)^{p-2}\left(\frac{\gamma_{\rm c}^s}{\hat{\gamma}_{\rm c}^s}\right)^{\frac{p-3}{2}}\right)^\frac{2}{p+2}$ \\
&$ \gamma_{m} \ll \gamma_{c}$ &$\hat{\gamma}_{m} < \gamma_{\rm c} < \hat{\gamma}_{c}$ &$ Y \ll 1$ &$Y(\nu_{c})=\frac{\epsilon_{\rm e}}{\epsilon_{\rm B}(3-p)}\left(\frac{\gamma_{\rm m}}{\gamma_{\rm c}^s}\right)^{p-2}\left(\frac{\gamma_{\rm c}^s}{\hat{\gamma}_{\rm c}^s}\right)^{\frac{p-3}{2}}$ \\
&$ \gamma_{m} \ll \gamma_{c}$ &$\hat{\gamma}_{m} < \gamma_{\rm c}$&$Y \gg 1$&$Y(\nu_{c}) = \left(\frac{\epsilon_{\rm e}}{\epsilon_{\rm B} (3-p)}\left(\frac{\gamma_{m}}{\hat{\gamma}_{m}}\right)^{-\frac{4}{3}}\left(\frac{\gamma_{m}}{\hat{\gamma}_{c}^s}\right)^{\frac{7}{3}}\right)^{\frac{3}{7}}$\\
&$ \gamma_{m} \ll \gamma_{c}$ &$\hat{\gamma}_{m} < \gamma_{\rm c}$&$Y \ll 1$&$Y(\nu_{c}) = \frac{\epsilon_{\rm e}}{\epsilon_{\rm B} (3-p)}\left(\frac{\gamma_{m}}{\hat{\gamma}_{m}}\right)^{-\frac{4}{3}}\left(\frac{\gamma_{m}}{\hat{\gamma}_{c}^s}\right)^{\frac{7}{3}}$\\
\hline
\end{tabular}
\end{table*}

\begin{figure}
    \centering
    \includegraphics[width=0.47\textwidth]{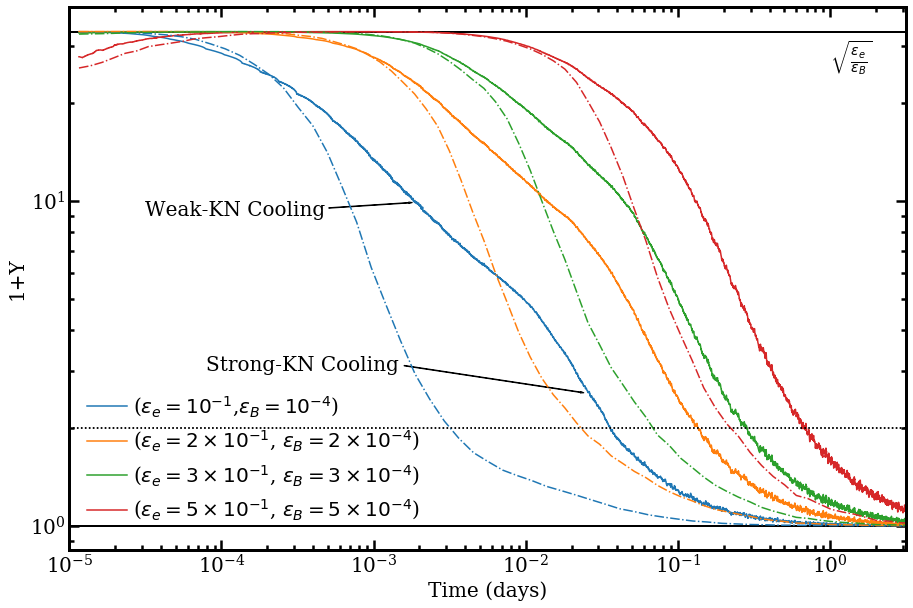}
    \caption{$Y$ as a function of time for typical X-ray (1~keV; solid) and high-energy gamma-ray (0.1~GeV; dashed) observing bands, for various values of $\epsilon_{\rm e}$ and $\epsilon_{\rm B}$ (and other parameters the same as in Figure~\ref{Yfull}). The dotted line indicates $1+Y=2$, below which SSC cooling is no longer dominant. The maximum $Y$ is dictated by $\sqrt{\epsilon_{\rm e}\epsilon_{\rm B}^{-1}}$ (upper solid black line), but the duration of the effect depends strongly on the individual values $\epsilon_{\rm e}$ and $\epsilon_{B}$. As $\epsilon_{\rm e}$ ($\epsilon_{\rm B}$) increases, the X-ray curve transitions from a shape dominated by $Y_{\rm T}$, denoted by the second break during the decay phase, to KN-cooling dominated at $\epsilon_{\rm e}=0.5$ ($\epsilon_{\rm B}=5\times 10^{-4}$). The high-energy gamma-ray curves are always strongly KN suppressed once they drop below the maximum.} 
    \label{KNfreq}
\end{figure}

\subsection{Implementing KN Effects: $Y$ at $\nu_{\rm c}$}

Unlike the Thomson case, $Y$ has a strong $\gamma_{\rm e}$ dependence when KN-suppression is important. Because we are dealing with a code for fitting observational data, it makes sense to continue discussing $Y$ as $Y(\nu_{\rm e})$. With that in mind, we need to have a solution specific to $\nu_{\rm c}$, so that we can properly determine the cooling regime and the location of $\gamma_{\rm c}$-defined breaks. To do this, we need to solve our KN-approximated solution for $\gamma_{\rm e}=\gamma_{\rm c}$, and produce a solution that is agnostic to both the cooling regime and the KN regime at $\nu_{\rm c}$. We again invoke a smoothly broken power-law approximation, combining not just the fast- and slow-cooling regimes, but also the weak and strong KN regimes of each solution. This results in a nested series of smoothly broken power-law solutions that culminate in an approximate description of $Y\left(\nu_{\rm c}\right)$ in all regimes. With a continuous solution for $Y\left(\nu_{\rm c}\right)$, we can determine the cooling regime and calculate $Y\left(\nu_{\rm c}\right)$ using the power law functions defined in Section~\ref{KNsec}; these solutions are found in Table~\ref{tab:Yckn}. We plot $Y(\nu_{\rm e})$ as a function of time for several frequencies and parameters in Figure~\ref{KNfreq}. In each regime, we see the breaks due to KN effects, and we extract an additional break defined as $\gamma_0$ in \citet{Nakar} at $Y=1$.

\subsection{Computational Complexity}

Because \texttt{boxfit} allows for multiple emission times and regions to be taken into account, even at a single observer time, we cannot define a global $Y(\nu_{\rm c})$ and must calculate it for every emitting point in the jet. This does add to the computational complexity and time required to run, with a Thomson-solution-enabled version of \texttt{boxfit} running about 20-70\% slower than the comparable synchrotron-only version, depending heavily on the simulation resolution and the observer angle. In realistic and typical examples, this has caused the fit time to increase by about 40\% when none of the fit parameters are fixed.  The KN-enabled version presents additional hurdles as the $Y(\nu_{\rm c})$ parameter becomes more complex in this case, and there is an additional calculation that includes a series of Boolean checks for every grid point. The overall effect on run time still remains within 50\%. SSC effects can be enabled at compile time using the variables in the \texttt{environment} header file both with and without KN effects.

As an alternative to the approach described in this paper, $Y$ could be solved for a grid of values and tabulated for use in \texttt{boxfit} in a similar manner to how the jet dynamics are included. This would be straightforward in the Thomson regime, as $Y$ only explicitly depends on three parameters $\left(\epsilon_e\epsilon_B^{-1},\ \gamma_{\rm m}\gamma_{\rm c}^{-1},\ p\right)$. Providing a sufficient sample to characterize the behavior of $Y$ in the KN regime would likely be more difficult as the behavior is more complex and dependent on the individual values of the above parameters, along with $\gamma_e$, in addition to their ratios.

\section{Effects on Broadband Spectra and Light Curves}\label{effs}

\begin{figure}
    \includegraphics[width=0.47\textwidth]{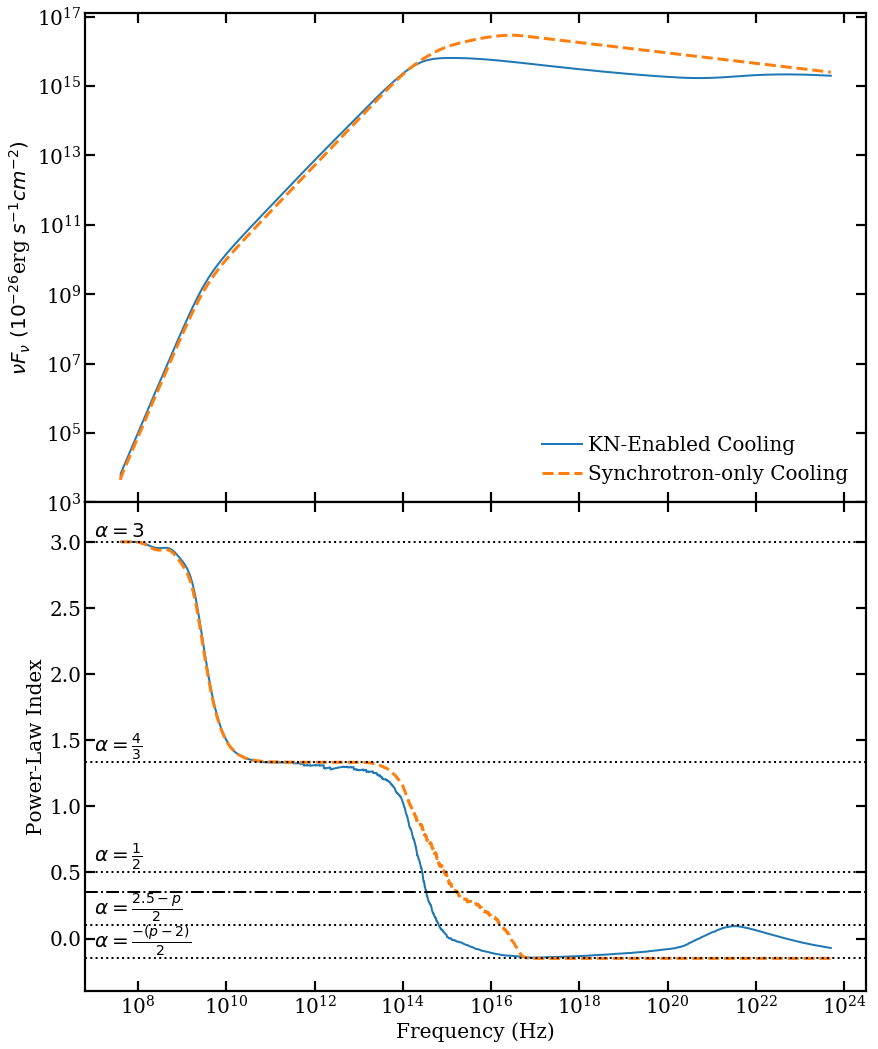}
    \caption{Example of a fast-cooling energy spectrum (upper panel) and spectral power-law indices (lower panel). This and the slow-cooling case are the most common instantaneous spectra produced with typical parameters, but other spectra may play important transitional roles. The power-law indices of the two energy spectra differentiate between the spectral cases. The dotted lines correspond to slopes we would expect from a fast-cooling spectrum exhibiting SSC cooling with KN effects. The $\frac{1}{2}$ slope is absent from the KN spectrum because the afterglow is nearing the transition from fast to slow cooling. This becomes readily apparent when compared to the synchrotron spectrum, which has already entered slow-cooling and shows a new break beginning to form with a slope of $\frac{3-p}{2}$ (thick dash-dotted line).}
    \label{fig:nI}
\end{figure}

\citet{Nakar} present the effects of SSC on the synchrotron spectrum in great detail, and our aim is to import their results into a framework where the mathematics are agnostic to the cooling and KN regime. Additionally, broadband modeling with \texttt{boxfit} is more sensitive to the time evolution in the data set of a given GRB afterglow, as opposed to determining the exact spectral regime at any given instant. Therefore, we focus largely on the evolution of afterglow light curves. We do present an example of a typical afterglow spectrum in Figure~\ref{fig:nI}.

\begin{figure}
    \includegraphics[width=0.465\textwidth]{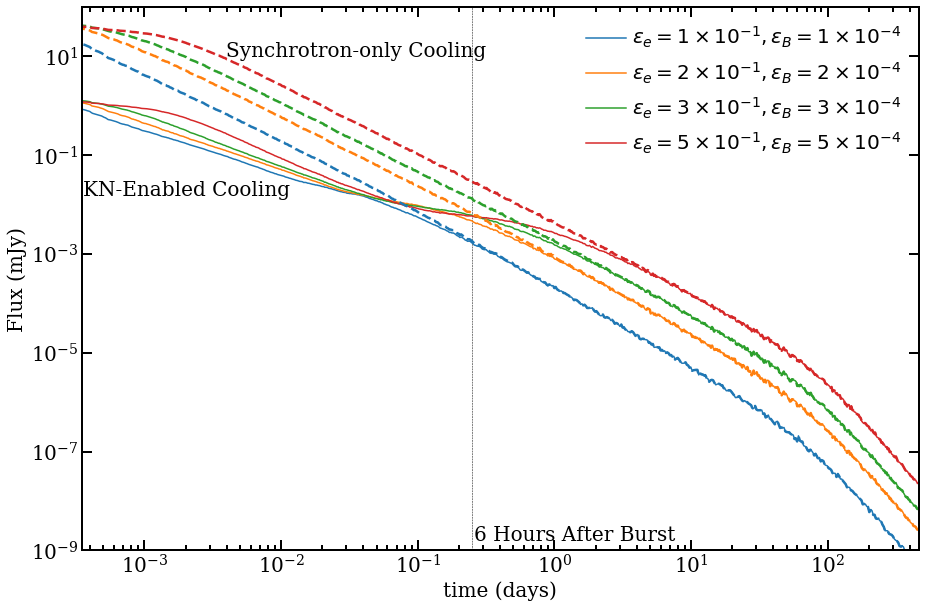}
    \caption{X-ray (1~keV) light curves corresponding to the $Y$ curves in Figure~\ref{KNfreq}, for various values of $\epsilon_{\rm e}$ and $\epsilon_{\rm B}$ (and other parameters the same as in Figure~\ref{Yfull}). The overall shifts in the light curves are due to the effect of $\epsilon_{\rm e}$ and $\epsilon_{\rm B}$ on the synchrotron spectrum. The strength of the suppression between the synchrotron baseline (dashed lines) and the solid curves are due to SSC effects. The transition time and behavior do vary significantly as we vary the two microphysical parameters. The vertical line indicates the approximate transition time between the Thomson and KN regimes.}
    \label{xrteeeb1e3}
\end{figure}

\begin{figure}
    \includegraphics[width=0.47\textwidth]{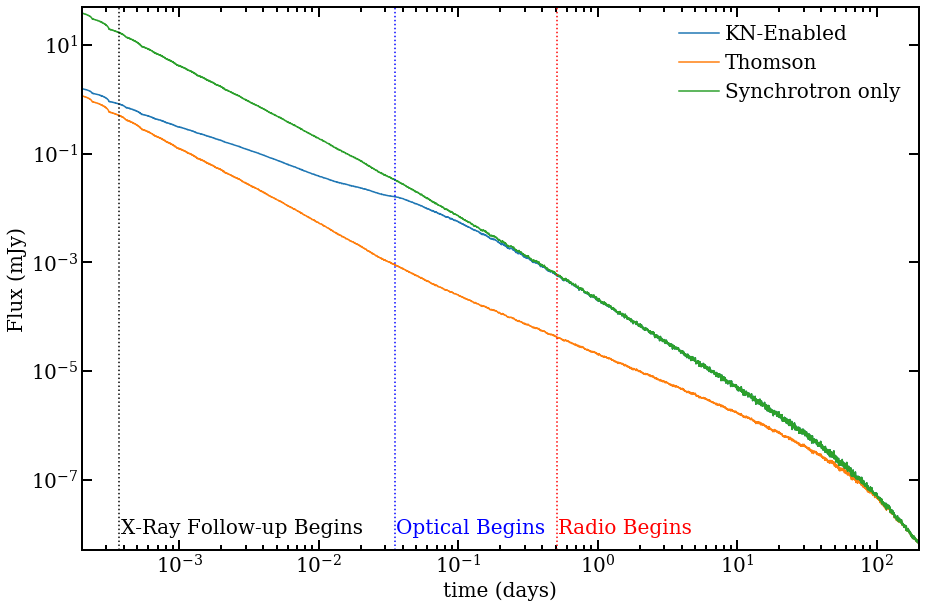}
    \caption{X-ray (1~keV) light curves for each of the possible cooling mechanism combinations using the parameters listed in Figure~\ref{Yfull} ($E_{\rm iso}=10^{53}$ergs, $p=2.5$, $\epsilon_{\rm e}=10^{-1}$, $\epsilon_{\rm B}=10^{-4}$, $\xi=1$, $\theta_0=0.3$ rad, $\theta_{\rm obs}=0.0$). With these fairly typical parameters, the over-estimation of SSC cooling caused by failing to include KN effects is evident. The vertical lines indicate typical times at which X-ray, optical and radio observations of GRB afterglows commence.}
    \label{xrt3}
\end{figure}

SSC, both with and without KN effects, can have a profound effect on light curve behavior depending on the microphysical parameters. Parameters such as the isotropic equivalent energy and density are degenerate in terms of their qualitative behavior, and differences from varying one or the other do not provide distinct changes in how $Y$ varies. The parameters $\epsilon_{\rm e}$ and $\epsilon_{\rm B}$ have a more direct impact on the strength and longevity of SSC effects, which makes sense given that $Y$ depends on powers of $\epsilon_{\rm e}\epsilon_{\rm B}^{-1}$. One would naively expect a direct relationship between this ratio and the magnitude of any SSC effects, but this is an incomplete description of reality. KN effects provide a direct dependence on individual values of $\nu_{\rm c}$ and $\nu_{\rm m}$ as opposed to only their ratio, which introduces a direct dependence on $\epsilon_{\rm e}$ and $\epsilon_{\rm B}$ as individual parameters. This results in different behavior in $Y$ that would not appear when considering only scattering in the Thomson regime. Figure~\ref{KNfreq} shows how varying $\epsilon_{\rm B}$ affects the flux suppression, with $Y$ increasing up until $\epsilon_{\rm B} \sim 10^{-4}$, beyond which KN effects dominate. There is a distinct set of breaks in the light curves caused by the addition of KN effects, as illustrated in Figure~\ref{xrteeeb1e3}. Inverse-Compton cooling on its own does not show obvious breaks, and presents an overall suppression of the flux, followed by a smooth and largely continuous rise back to the synchrotron light curve at late times. For most parameters, $Y$ is significantly overestimated, especially at later times when KN effects drive $Y$ much more quickly to 0 than would be predicted by the $Y$ computed in their absence, as seen in Figure~\ref{xrt3}.

\begin{figure}
    \includegraphics[width=0.47\textwidth]{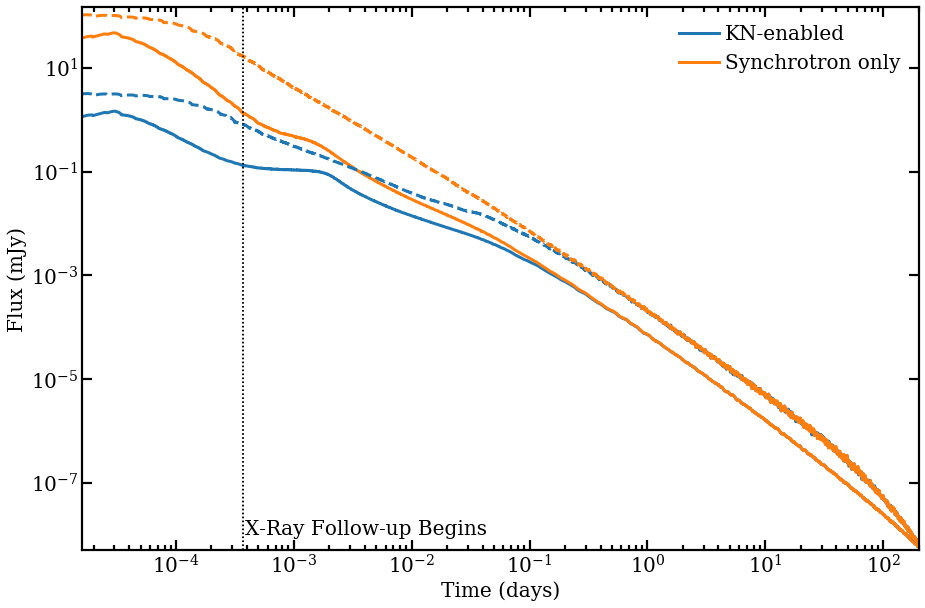}
    \caption{X-ray (1~keV) light curves for the parameters listed in Figure~\ref{Yfull} (dashed; $E_{\rm iso}=10^{53}$ergs, $p=2.5$, $\epsilon_{\rm e}=10^{-1}$, $\epsilon_{\rm B}=10^{-4}$, $\xi=1$, $\theta_0=0.3$ rad, $\theta_{\rm obs}=0.0$), together with curves for the same parameters except for $\theta_{\rm obs}=\theta_0=0.3$ rad (solid). The quantitative effects of the observer angle on $Y$ can be seen in Figure~\ref{fig:y-off}. The vertical line indicates the typical time scale at which X-ray observations of GRB afterglows commence.}
    \label{xrtonoff}
\end{figure}

\begin{figure}
    \includegraphics[width=0.45\textwidth]{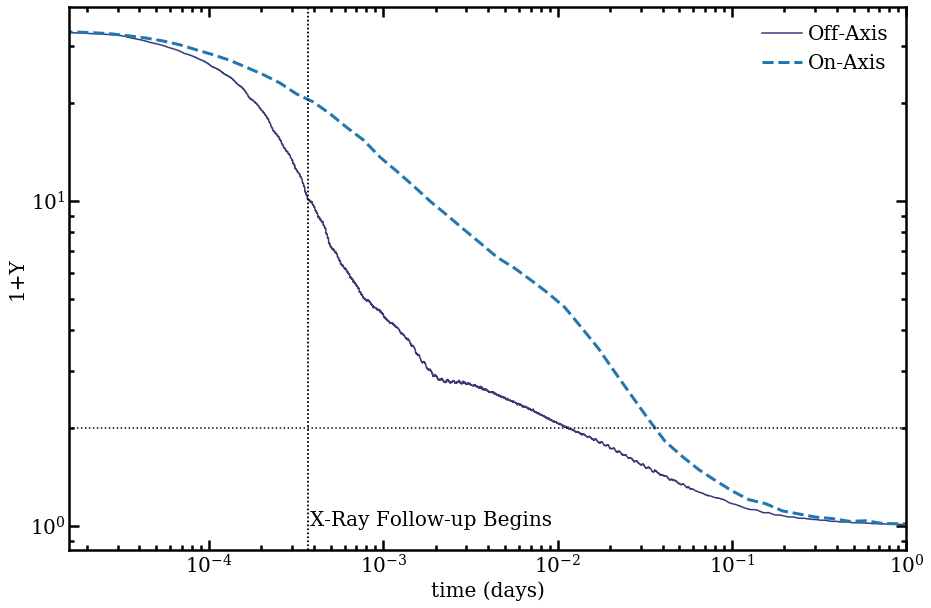}
    \caption{The observed $Y$ parameter as defined by the ratio of the synchrotron to SSC power for the light curves in Figure~\ref{xrtonoff}. Once both values leave the early $Y_{\rm T}$ behavior, $Y_{\text{on}}$ shows stronger early time suppression, while $Y_\text{{off}}$ exhibits lower suppression for a longer period of time. The vertical line indicates the typical time scale at which X-ray observations of GRB afterglows commence.}
    \label{fig:y-off}
\end{figure}

\begin{figure}
    \includegraphics[width=0.45\textwidth]{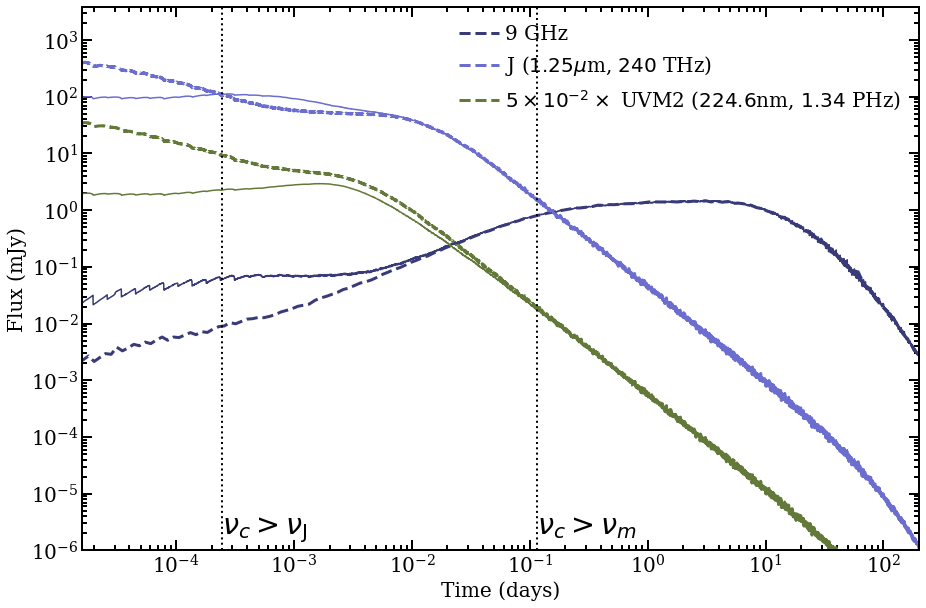}
    \caption{SSC cooling effects (solid lines) on spectral bands in the radio (9 GHz), near-infrared (J) and ultraviolet (UVM2) compared to the synchrotron-only cooling for the same bands (dashed), for the same parameters as Figure~\ref{Yfull}. The J band is unique for this simulated set of parameters, in that it initially exhibits suppression, followed by a re-brightening as $\nu_{\rm c}$ passes through the observing band, indicated by the left-most vertical line.}
    \label{fig:bump-spec}
\end{figure}

Most of the other parameters ($E_{\rm iso}$, $\theta_0$, $n_0$ and $\xi$) result in similar variations to the light curves as modifying $\epsilon_{\rm e}$, since those parameters modify the total energy available to the electron population. $\epsilon_{\rm B}$ is an exception, since it dictates the transition time between $Y_{\rm T}$ and $Y_{\rm KN}$, while $\epsilon_{\rm e}$ drives the magnitude of $Y$ and has a strong impact on the power-law index of the light curve as it transitions back to synchrotron-only cooling. $\theta_{0}$ also effects the jet-break time, but that occurs well beyond the end of observed SSC influence for all parameters. The observer angle $\theta_{\rm obs}$ requires a little more attention: increasing the observer angle results in lower emission at earlier times, which would seem to be an issue for detecting SSC cooling in the afterglow. A larger observer angle also means that emission from the far edge of the jet will be arriving at a later time than the same emission on the near side of the jet. This results in an initial increase in flux after detection as more of the jet becomes visible, as well as changes to the decay of the light curves. One of the biggest changes is a significantly smoothed and chromatic jet break. The $Y$ parameter in an off-axis jet becomes more complicated as the jet now contains observed cooling asymmetries. These asymmetries, combined with SSC cooling, can result in structures such as plateaus and re-brightening events which appear in the light curves in Figure~\ref{xrtonoff} \citep[see also e.g.][]{benpl1,benpl2}. $Y$ shows a more rapid decay initially when viewed off-axis, but takes longer to reach $Y=1$ than in the on-axis case, as can be seen in Figure~\ref{fig:y-off}.

A final change to light curve behavior we discuss here occurs for frequencies $\nu < \nu_{c}$. In the case of fast cooling, the synchrotron peak $F_{\nu}(\nu_{\rm c})$ is pushed to lower frequencies, resulting in an increase in flux compared to the synchrotron-only case. For observations within this regime, we see a marked increase in flux that transitions to the synchrotron-only behavior as $\nu_{\rm c}$ approaches $\nu_{\rm m}$. We present an on-axis example in the next section for completeness, but note that such observations would have to occur within minutes of a burst being detected. Likewise, we could potentially see suppression or increased emission in the optical band, depending on the location of $\nu_{\rm c}$, but this would also require very early observations to detect. The exact behavior of three example bands in the radio, near-infrared and ultraviolet are presented in Figure~\ref{fig:bump-spec}.

\section{SSC Effects on Microphysical Parameters from Model Fitting}\label{mod}

The results presented in the previous section indicate that including SSC effects can have a significant impact on the light curves in various wavebands. Therefore, not including SSC effects in modeling of broadband data sets may result in a misinterpretation of the characteristic spectral breaks, in particular $\nu_{\rm c}$; and as a result, the physical parameters derived from the spectral breaks may be significantly off from the true values. To quantify the changes to physical parameters based on afterglow model fitting, we simulated two afterglows based on the on- and off-axis cases discussed in the previous section. We generated synthetic light curves at various wavelengths covering the radio, millimeter, near-infrared, ultraviolet and X-ray regimes, using our new implementation of \texttt{boxfit} with SSC and KN effects enabled. The light curves were sampled with cadences that are fairly typical of currently available instrumentation, and the light curve start times are consistent with the vertical lines in Figure~\ref{xrt3}. Gaussian noise was added to each data point with errors consistent with observed bursts with similar fluxes in the respective wave bands. We then performed iterative fitting using \texttt{boxfit} with and without SSC and KN effects enabled. We did not consider the case with $Y=Y_{\rm T}$, as the light curves in that case appear to be far more suppressed than would be expected in reality. We include a selection of the fit light curves that showcase the difference in fitting that results from attempting to fit the KN enabled synthetic afterglows with a model that only includes synchrotron cooling below. We also discuss the resulting changes in derived parameters in each case.

\subsection{On-Axis ($\theta_{\rm obs}=0$)}

\begin{figure*}
    \centering
    \includegraphics[width=0.98\textwidth]{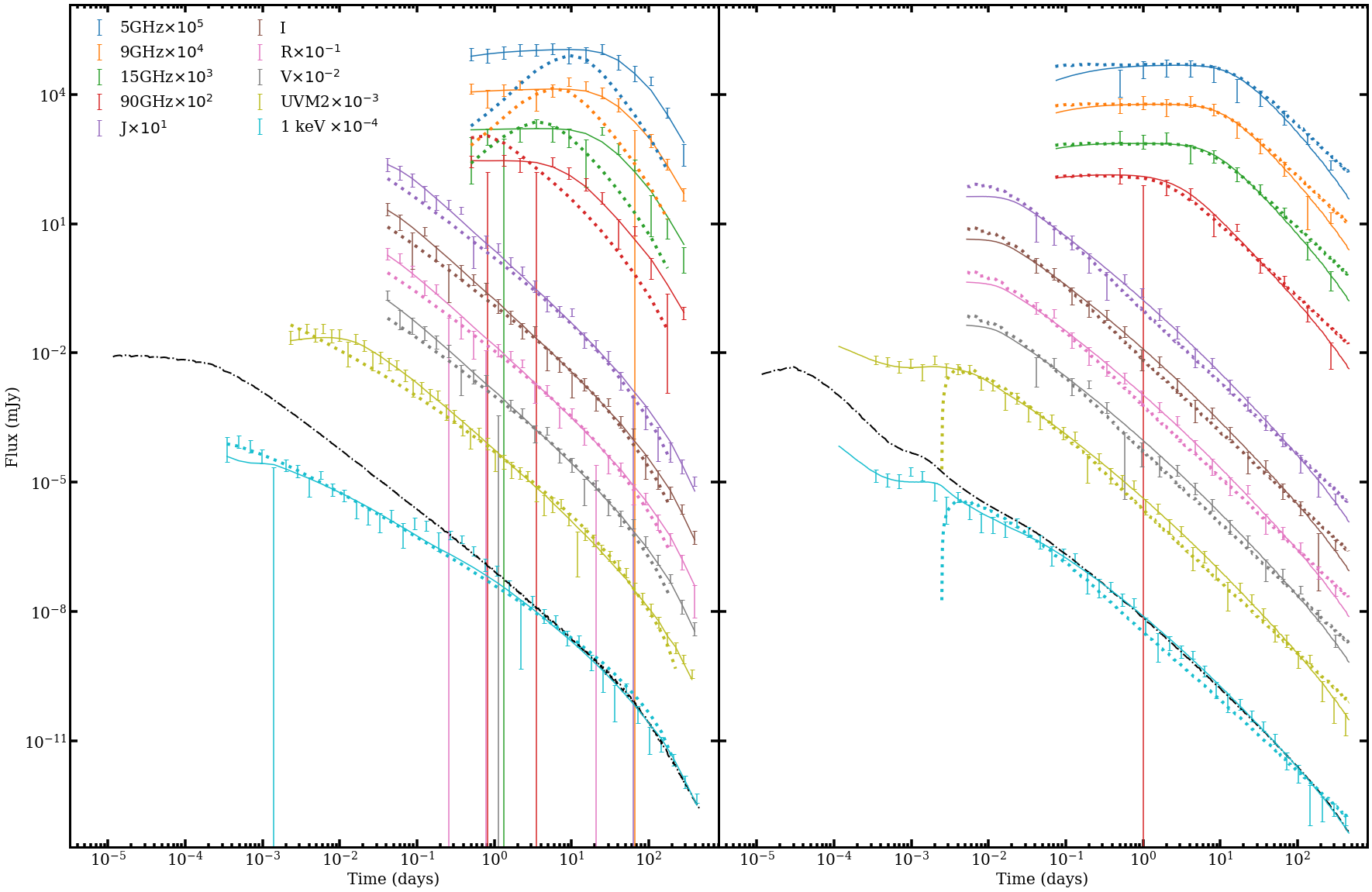}
    \caption{Broadband light curve fits for our simulated on-axis (left) and off-axis (right) data sets. The SSC/KN-Cooling fits are solid lines, while the synchrotron-only fits are the dashed ones. The black dash-dotted curve is a synchrotron-only X-ray light curve generated from the simulated input parameters. Each band has been multiplied by a factor for ease of readability (see the legend in the top left corner for the multiplication factors for each band).}
    \label{fig:on_fits}
\end{figure*}

\begin{figure*}
    \centering
    \includegraphics[width=0.7\textwidth]{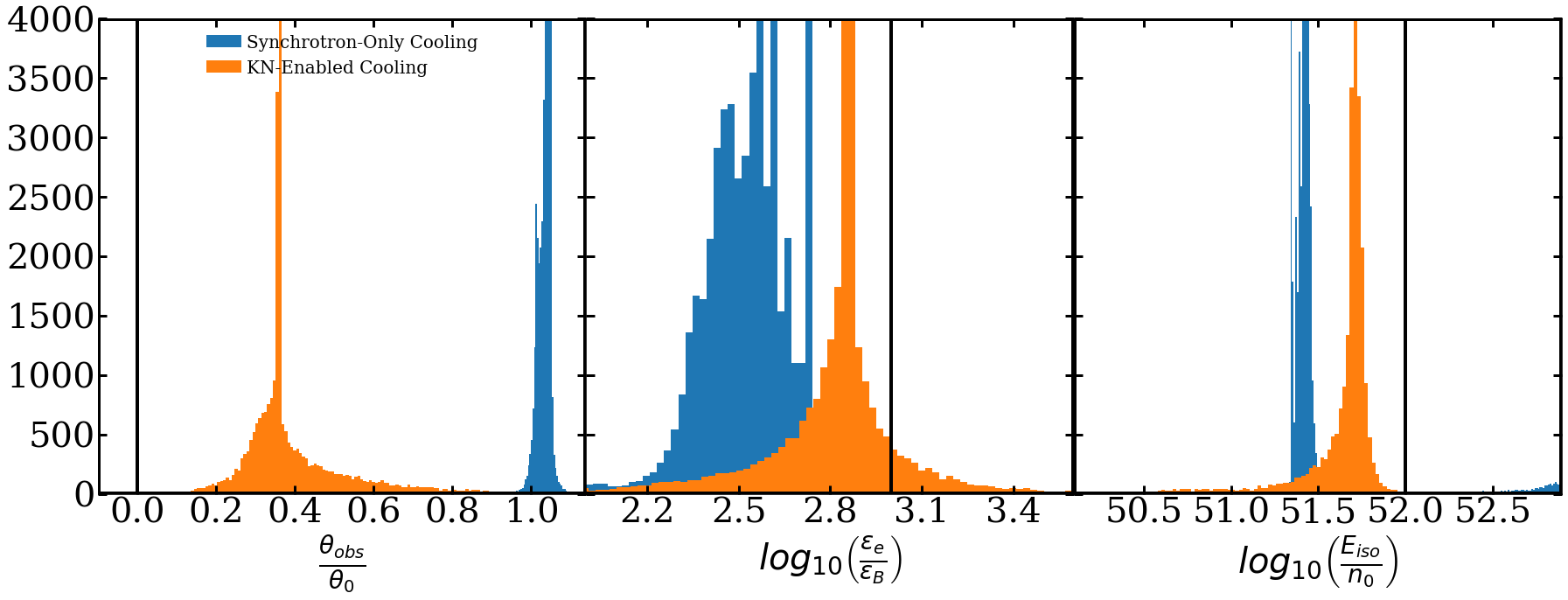}
    \caption{Histograms for non-degenerate parameter ratios for the on-axis fits in Figure~\ref{fig:on_fits}: $\theta_{\rm obs}/\theta_{0}$ (left panel), $\log_{10}\left(\epsilon_{\rm e}/\epsilon_{\rm B}\right)$ (middle panel), and $\log_{10}\left(\rm{E}_{\text{iso}}/n_{0}\right)$. The black lines indicate the input values for the simulated light curves, the blue histograms are for synchrotron-only and the orange histograms for SSC/KN-Cooling fits.}
    \label{fig:logeeeb}
\end{figure*}


\begin{table*}
\caption{Model Fit Parameters for an On-Axis ($\theta_{\rm obs}=0$) Afterglow}
\label{tab:on-par}
\begin{tabular}{lcccccccll}
\hline
 &$\theta_0$&E${}_{\text{iso}}(10^{53}$ \text{ergs})&$n_{0}$&$\theta_{\rm obs}$&$p$&$\epsilon_{\rm B} (10^{-4})$&$\epsilon_{\rm e}$&$\xi$&$\chi^2_{\text{red}}$\\
\hline
Input&$0.3$&$1$&$1$&$0$&$2.5$&$3$&$0.3$&$1$&...\\
SSC/KN-Cooling&$0.43^{+0.04}_{-0.01}$&$0.62^{+0.09}_{-0.28}$&$12^{+2}_{-4}$&$0.15^{+0.08}_{-0.05}$&$2.49^{+0.01}_{-0.06}$&$2.9^{+4.2}_{-0.7}$&$0.21^{+0.02}_{-0.03}$&$1$& $1.2$\\
Synchrotron&$0.20^{+0.01}_{-0.02}$&$6.8^{+1.2}_{-2.8}$&$260^{+100}_{-100}$&$0.207^{+0.004}_{-0.004}$&$2.01^{+0.16}_{-0.01^*}$&$0.79^{+0.62}_{-0.33}$&$0.023^{+0.003}_{-0.005}$&$1$&$4.2$\\
\hline
\end{tabular}
\end{table*}

The on-axis case is the more straightforward of the two cases, and the full fits are shown in the left-hand panel of Figure~\ref{fig:on_fits}. Both models, with and without SSC/KN-Cooling, give a reasonable fit to the synthetic X-ray, ultraviolet, optical and near-infrared light curves. The major issues in light curve reconstruction occur when simultaneously fitting the X-ray and radio bands, which has a profound impact on the other observed bands. The simulated X-ray light curve is the only one that shows obvious signatures of SSC cooling, with early time suppression and an extended flattening of the light curve during the transition back to the simulated synchrotron curve, which happens at about a day after the burst. Figure~\ref{fig:on_fits} shows the X-ray band fit plotted together with the X-ray light curve for a synchrotron-only model with the simulated parameters. The differences in fits not only explain the need for including KN effects in any afterglow model, but it also demonstrates the need for having robust radio data for modeling broadband afterglows.

In order to match the decreased early time emission, the synchrotron curve requires significant changes to the derived physical parameters, up to more than an order of magnitude (see Table~\ref{tab:on-par} and  Figure~\ref{fig:logeeeb}), and no longer resembles the unmodified synchrotron curve. These changes result in significantly less emission at early times, but unlike the SSC/KN case, the synchrotron emission is not being up-scattered to higher energies. Instead, the overall emission of the afterglow is lower, causing the radio curves to appear significantly less bright than they should. Additionally, the synchrotron curve also required the observing angle to be larger than the opening angle of the jet, with $\theta_{\rm obs}/\theta_{0}$ being approximately 1, far from the simulated value, as can also be seen in Table~\ref{tab:on-par} and  Figure~\ref{fig:logeeeb}.

\subsection{Off-Axis ($\theta_{\rm obs}=\theta_0$)}

\begin{figure*}
    \centering
    \includegraphics[width=0.7\textwidth]{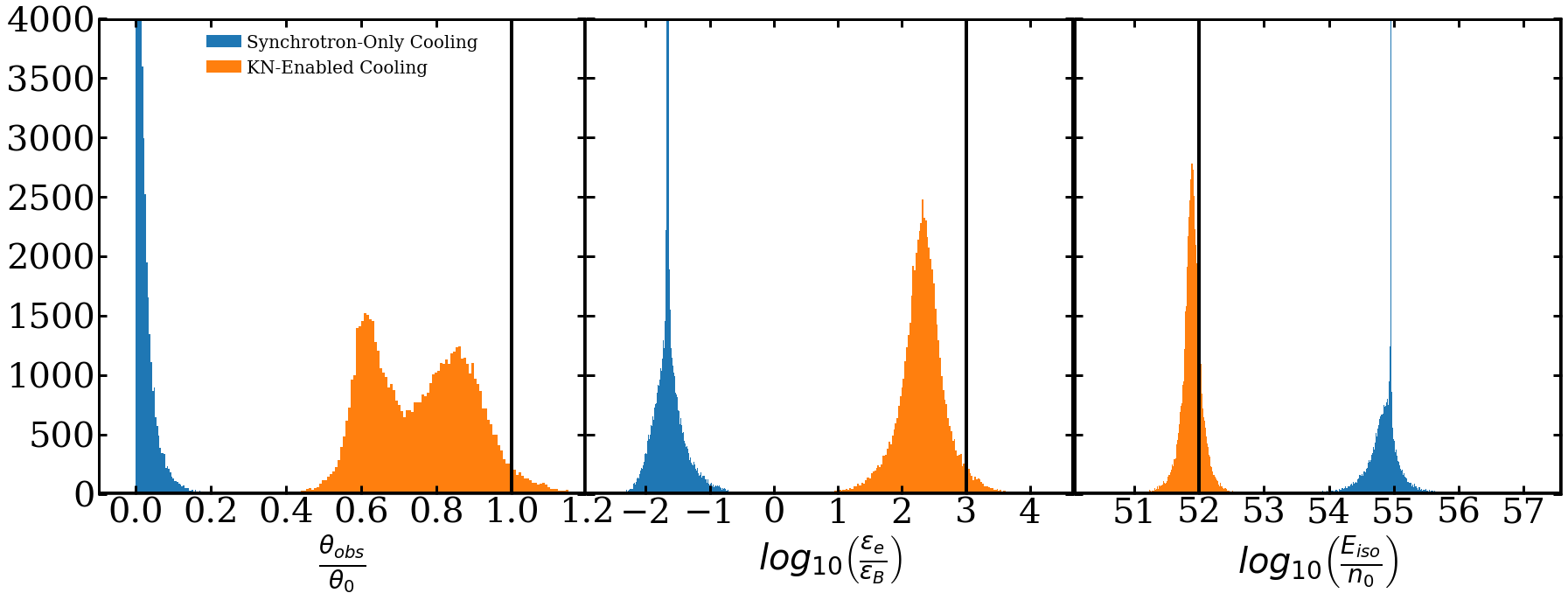}
    \caption{Histograms for non-degenerate parameter ratios for the off-axis fits in Figure~\ref{fig:on_fits}: $\theta_{\rm obs}/\theta_{0}$ (left panel), $\log_{10}\left(\epsilon_{\rm e}/\epsilon_{\rm B}\right)$ (middle panel), and $\log_{10}\left(\rm{E}_{\text{iso}}/n_{0}\right)$. The black lines indicate the input values for the simulated light curves, the blue histograms are for synchrotron-only and the orange histograms for SSC/KN-Cooling fits.}
    \label{fig:logeeeb-off}
\end{figure*}

Our off-axis modeling exhibits similar results to those of the on-axis case, and the full fits are shown in the right-hand panel of Figure~\ref{fig:on_fits}. The synchrotron fits misinterpret several key parameters in an attempt to compensate for the missing breaks resulting from SSC effects, for some parameters up to more than three orders of magnitude (see Table~\ref{tab:off-par} and  Figure~\ref{fig:logeeeb-off}). The clearest result of this is an inversion in values of $\epsilon_{\rm e}$ and $\epsilon_{\rm B}$, along with an observer angle that indicates a nearly on-axis observer. Unsurprisingly, SSC cooling does a much better job at constraining the fit parameters, even though it does struggle with the observer angle, producing a bi-modal parameter distribution just below the simulated value. This is in part due to the opening angle being wider than simulated, resulting in a similar $E_{\text{jet}}$ in spite of the smaller value of $E_{\text{iso}}$. The values of  $\epsilon_{\rm e}$ and $\epsilon_{\rm B}$ are both well recovered, as are non-degenerate quantities such as $E_{\rm iso}n_0^{-1}$ (Figure~\ref{fig:logeeeb-off}). In general, $\epsilon_{\rm B}$ seems harder to constrain than $\epsilon_{\rm e}$ which may be caused by how well constrained $\epsilon_{\rm e}$ is by the radio band \citep{VB1}. Freezing the observer angle did allow for a minor increase in accuracy in recovering $\frac{\epsilon_{e}}{\epsilon_B}$, with the ultimate limitation being the uncertainty in the X-ray lightcurve near the transition back to synchrotron-dominated cooling.

\begin{table*}
\caption{Model Fit Parameters for an Off-Axis $\left(\theta_{\rm obs}=\theta_0\right)$ Afterglow}
\label{tab:off-par}
\begin{tabular}{lcccccccll}
\hline
 &$\theta_0$&E${}_{\text{iso}}(10^{53}$ \text{ergs})&$n_{0}$&$\theta_{\rm obs}$&$p$&$\epsilon_{\rm B} (10^{-4})$&$\epsilon_{\rm e}$&$\xi$&$\chi^2_{\text{red}}$\\
\hline
Input&$0.3$&$1$&$10$&$0.3$&$2.5$&$1$&$0.1$&$1$&...\\
SSC/KN-Cooling&$0.48^{+0.02}_{-0.08}$&$0.30^{+0.13}_{-0.08}$&$3.6^{+1.9}_{-0.9}$&$0.30^{+0.06}_{-0.02}$&$2.49^{+0.03}_{-0.04}$&$4.1^{4.2}_{-1.7}$&$0.10^{+0.01}_{-0.01}$&$1$&$0.82$\\
Synchrotron&$0.42^{+0.02}_{-0.03}$&$2.9^{+3.3}_{-1.9}$&$0.033^{+0.021}_{-0.011}$&$0.0002^{+0.0023}_{-7\times10^{-5}}$&$2.55^{+0.05}_{-0.01}$&$3400^{+3200}_{-2200}$&$0.0076^{+0.0041}_{-0.0024}$&$1$&$2.6$\\
\hline
\end{tabular}
\end{table*}

\subsection{Potential Effects on The Observed GRB Parameter Distribution}

While some of the physical parameters derived from the synchrotron-only fits are somewhat unusual, and in some cases orders of magnitude away from the simulation input parameters, they do not appear non-physical. It would be natural for someone performing a fit without prior knowledge of the physical parameters to assume the burst is well constrained by their synchrotron-only model, despite the fact that the spectrum was created by significantly different physical parameters and in a regime where SSC cooling is in fact important. This demonstrates that systematic biases in the values of inferred parameters from modeling can arise when SSC is not treated self-consistently. It is worth noting that the better fits including SSC/KN-Cooling are mainly due to the early X-ray data points, highlighting the need for such observations. That fact, coupled with the poor radio fits in the on-axis case, point to a need to re-evaluate our current understanding of the underlying parameter distributions derived from modeling with synchrotron-only models.

At a population level, \citet{BNP16} have shown that not accounting for SSC cooling effects leads to artificially enhanced values of the GRB prompt efficiency and its scatter, which are also inconsistent with independent constraints from Fermi-LAT detected GRBs \citep{Nava}. For individual GRBs, these effects have been illustrated by GRBs with well-constrained radio data for which broadband modeling has been a challenge \citep{Gv14}. In those cases, poor fits may be due to the modeling being largely constrained by the X-ray data. When including SSC effects, the X-ray light curve deviates significantly from what would be expected in a synchrotron-only model, and any attempt at fitting the X-rays well will result in parameters that do not represent the underlying physics. The optical fits are less sensitive to these variations in the X-rays, but the radio data, which are strongly influenced by deviations in $\epsilon_{\rm B}$ and $\epsilon_{\rm e}$, are affected significantly. For bursts lacking early-time X-ray observations, SSC effects can still alter the X-ray light curve up to $\sim 1$ day or more depending on the parameters, and may only become noticeable when a broadband fit is performed. There are ways to produce a good radio fit with unusual, but still physical, parameters. This can be seen clearly in the off-axis fits where the synchrotron-only fit works well at late times, but fails to fit the early time X-ray data, highlighting the need for broadband coverage of the afterglow over a long time span for successful model fitting.

Population studies based purely on X-ray data will be particularly affected, because it is still possible to get a well-constrained X-ray fit even with a synchrotron-only model. Parameters that would not produce large SSC cooling effects are still going to be reasonable, but single band fits are likely not well constrained in general. The best course of action would be to examine a large sample of bursts with early-time X-ray observations, coupled with well-sampled optical and radio light curves. Such recommendations are not new, but including SSC effects drives home the fact that X-ray and optical data alone are not sufficient to constrain the (micro)physical parameters of the afterglow. 
    
Finally, we note that all modeling performed using \texttt{boxfit} and related numerical techniques still contain certain systematic biases. In particular, the global treatment of cooling in \texttt{boxfit} leads to an underestimate of flux above the cooling break \citep{vezmf10}. This effect should have only a minor impact on parameter comparisons as our modifications would be subject to the same systematic uncertainties as a synchrotron model with identical parameters. The effects will be important when performing iterative fitting on observational data and will need to be considered in the same fashion as for the original \texttt{boxfit}.

\section{Summary and Future Work}\label{sum}

We have presented a methodology and implementation for fitting synchrotron and SSC cooling in broadband light curves from GRB afterglows based on the afterglow modeling package \texttt{boxfit} \citep{vetajvdh}. SSC effects were implemented based on a framework laid out in \citet{Nakar}, with modifications to remove any need to know the cooling and/or KN regime in advance. We applied these modifications to simulated data sets, to examine how they would change the derived physical parameters compared to a synchrotron-only model. We found significant differences between simulated versus extracted parameters in the synchrotron-only model, up to three orders of magnitude for some parameters. Finally, we discussed the impact these changes may have on previous GRB parameter studies, and stress the need for broadband modeling including both radio and early time X-ray data, in assessing the underlying physics, especially the microphysical parameters. 

Next steps would include re-examining the observed GRB parameter space by applying this technique to a sample of afterglows. The sample would be composed of afterglows which were well sampled in the radio bands and include early-time X-ray data. This will limit the total number of bursts available to the sample, but should result in well-defined constraints on the fit parameters. The model can also be extended to include the effects of SSC emission rather than just cooling, allowing broadband fits to include GeV and TeV emssion for bursts such as GRB 180720B \citep{HESS}, GRB 190114C \citep{MAGIC} and GRB 190829A, to be included in the fitting process \citep[e.g.][]{derishev,fraija,wang}. 
Another application of this effort will be to study (off-axis) afterglows of future short GRBs detected due to a gravitational wave (GW) trigger. These have so far been studied analytically \citep[e.g.][]{WuGW, FongGW,FraijaGW,BGG20} and numerically \citep[e.g.][]{WuGW2,GottliebGW,GillGW,LBM20}, but without taking into account SSC cooling. As discussed in the present work, this is well motivated so long as the observed bands lie below the cooling frequency at all times. Indeed the latter condition appears to be satisfied in GRB 170817. In future GW-detected GRBs this may no longer be the case, and the modeling developed here will become relevant.

\section*{Acknowledgements}
The authors would like to thank H.J. van Eerten for useful suggestions on several aspects of the work presented here. TEJ would also like to thank M.J. Moss and S.I. Chastain for discussions on the theoretical framework. TEJ acknowledges support from NASA Astrophysical Theory Program \#80NSSC18K0566. He additionally acknowledges support from the Chandra X-ray Center, which is operated by the Smithsonian Institution under NASA contract NAS8-03060. Some of the computations in this paper were conducted on the George Washington University High Performance Computing Cluster, Colonial One, and on the Smithsonian High Performance Cluster (SI/HPC), Smithsonian Institution (\url{https://doi.org/10.25572/SIHPC}). The research of PB was funded by the Gordon and Betty Moore Foundation through Grant GBMF5076.
This work made use of the following software packages: MatPlotLib \citep{Hunter:2007}, NASA ADS, NumPy \citep{van2011numpy},Pandas \citep{pandas}.

\bibliographystyle{mnras}

\onecolumn
\appendix

\section{Derivation of SSC in the Thomson Regime}\label{appT}

A single population of electrons is generating both the photon field and the scattered photon field, so we can modify the spectrum of synchrotron radiation through the use of $Y$ as defined in \citet{RL}:
\begin{equation}
\label{Y:gammal}
Y = \frac{4}{3}n_0^{\prime} \sigma_{\rm T} \Delta R^{\prime} \langle\gamma_{\rm e}^2\rangle
\end{equation}
Here $\sigma_{\rm T}$ is the Thomson scattering cross-section, $n_0$ is the electron number density, $\Delta R$ is the length of a thin emitting shell at the shock boundary, and $\langle \gamma_{e}^2\rangle$ is the second moment of the electron Lorentz factor distribution:
\begin{equation}
\langle\gamma_{\rm e}^2\rangle =\frac{1}{n_0^{\prime}} \int_{1}^{\infty}d\gamma_{\rm e} \frac{d n_0^{\prime}}{d\gamma_{\rm e}}\gamma_{\rm e}^2
\end{equation}
Primed variables are defined in the co-moving frame of the jet. Note that we have replaced the simple power-law with the differential electron energy distribution because we need to consider how cooling changes the electron population in time.  Likewise,
\begin{equation}
\label{n0:gen}
n_0^{\prime} = \int_{1}^{\infty}d\gamma_{\rm e}\frac{dn_0^{\prime}}{d\gamma_{\rm e}}
\end{equation}

The $\gamma_{\rm e}^2$ term in the equation for $Y$ means that the energy radiated in SSC emission scales identically to synchrotron emission, so we would need to modify the electron cooling equation to demonstrate the effects on the spectrum.  The modification is already presented above, and we can see that $Y$ only effects electrons cooling quickly for the same reason synchrotron losses only affect the same group of electrons.

\subsection{Fast Cooling}

For the fast-cooling case we have the following electron energy distribution:
\begin{equation}
\label{ne:fast}
\frac{dn_0^{\prime}}{d\gamma_{\rm e}}= 
\begin{cases} 
C \left(\frac{\gamma_{\rm e}}{\gamma_{\rm c}}\right)^{-2} & \gamma_{\rm c} \leq \gamma_{\rm e}\leq\gamma_{\rm m} \\
C\left(\frac{\gamma_{\rm m}}{\gamma_{\rm c}}\right)^{-2}\left(\frac{\gamma_{\rm e}}{\gamma_{\rm m}}\right)^{-p-1} & \gamma_{\rm m} < \gamma_{\rm e}\\ 
\end{cases}
\end{equation}
Inserting Equation \ref{ne:fast} into \ref{n0:gen} gives us
\begin{equation}
\label{n0:fast}
n_0^{\prime} = C\gamma_{\rm c} \left(1-\frac{p-1}{p}\frac{\gamma_{\rm c}}{\gamma_{\rm m}}\right)
\end{equation}
and
\begin{equation}
\langle \gamma_{\rm e}^2 \rangle = (\gamma_{\rm m}\gamma_{\rm c})\left[\frac{p-1}{p-2}-\frac{\gamma_{\rm c}}{\gamma_{\rm m}}\right]\left(1-\frac{p-1}{p}\frac{\gamma_{\rm c}}{\gamma_{\rm m}}\right)^{-1}
\end{equation}
Combining the latter with Equation (A1), and using the definitions of $\gamma_{\rm m}$ and $ \gamma_{\rm c}$, we derive the full expression for $Y$ in the fast-cooling regime

\begin{equation}
\label{Yt:fast}
Y(1+Y) = \frac{(p-2)\epsilon_{\rm e}}{(p-1)\epsilon_{\rm B}}\left[\frac{p-1}{p-2}(1+Y)-\frac{\gamma_{\rm c}^s}{\gamma_{\rm m}}\right]\left((1+Y)-\frac{p-1}{p}\frac{\gamma_{\rm c}^s}{\gamma_{\rm m}}\right)^{-1}
\end{equation}

Here we have used the relation between $\gamma_{\rm c}$ and $\gamma_{\rm c}^{\rm S}$ to remove any implicit $Y$ dependence.  Rearranging this equation, we arrive at a cubic function of Y that can be solved analytically to produce one real analytic solution. While the full solution is rather lengthy, we get the expected asymptotic result in the limit that $\gamma_{\rm c}\ll\gamma_{\rm m}$. We also match with the transition value of $Y$ described below.

\subsection{Transition from Fast to Slow Cooling}

In the limit that $\gamma_{\rm c} = \gamma_{\rm m}\equiv\gamma_*$, we require that the fast- and slow-cooling solutions return the same result, and that this result matches with the expectation from modifying the electron distribution such that
\begin{equation}
\frac{dn_0^{\prime}}{d\gamma_{\rm e}}= \begin{cases} 
      C\left(\frac{\gamma_{\rm e}}{\gamma_*}\right)^{-p-1} & \gamma_* \leq \gamma_{\rm e} \\
  \end{cases}
\end{equation}
Performing similar calculations as above, we derive
\begin{equation}
\langle \gamma_{\rm e}^2 \rangle = (\gamma_{\rm m}\gamma_{\rm c})\left[\frac{p-1}{p-2}-1\right]\left(1-\frac{p-1}{p}\right)^{-1}
\end{equation}
and
\begin{equation}
Y(1+Y)=\frac{(p-2)\epsilon_{\rm e}}{(p-1)\epsilon_{\rm B}}\left[\frac{p-1}{p-2}-1\right]\left(1-\frac{p-1}{p}\right)^{-1}
\end{equation}
In both cases, we made use of the fact that the $\gamma_*=\gamma_{\rm m}=\gamma_{\rm c}$ to simplify $Y$.  This equation can also be solved analytically to yield
\begin{equation}
Y =\frac{1}{2}\left(\sqrt{1+\frac{4p}{(p-1)}\frac{\epsilon_{\rm e}}{\epsilon_{\rm B}}}-1\right)
\end{equation}
\subsection{Slow Cooling}

In the slow-cooling regime, $\gamma_{\rm c}$ and $\gamma_{\rm m}$ are reversed, such that only a small fraction of electrons are cooling on a timescale comparable to that of the dynamical time scale of the shock. As a result
\begin{equation}
\label{ne:slow}
\frac{dn_0^{\prime}}{d\gamma_{\rm e}}= \begin{cases} 
      C\left(\frac{\gamma_{\rm e}}{\gamma_{\rm m}}\right)^{-p} & \gamma_{\rm m} \leq \gamma_{\rm e}\leq\gamma_{\rm c} \\
     C\left(\frac{\gamma_{\rm c}}{\gamma_{\rm m}}\right)^{-p}\left(\frac{\gamma_{\rm e}}{\gamma_{\rm c}}\right)^{-p-1} & \gamma_{\rm c} < \gamma_{\rm e}\\
   \end{cases}
\end{equation}

Given this electron energy distribution, we derive
\begin{equation}
\label{n0:slow}
n_0^{\prime} = C\left[\frac{\gamma_{\rm m}^p\gamma_{\rm c}^{1-p}}{p(1-p)}+\frac{\gamma_{\rm m}}{p-1}\right]
\end{equation}
and
\begin{equation}
\langle \gamma_{\rm e}^2 \rangle = \left[\frac{\gamma_{\rm m}^p\gamma_{\rm c}^{1-p}}{p(1-p)}+\frac{\gamma_{\rm m}}{p-1}\right]^{-1}\left(\frac{\gamma_{\rm m}^3}{p-3}+\frac{\gamma_{\rm m}^{p}\gamma_{\rm c}^{3-p}}{(3-p)(p-2)}\right)
\end{equation}
From this, we find that
\begin{equation}
\label{Yt:slow}
Y(1+Y) = p\left[\frac{\epsilon_{\rm e}}{\epsilon_{\rm B}}\frac{\gamma_{\rm m}}{\gamma_{\rm c}^s}(1+Y)^{p-1}\frac{p-2}{p-3}+\frac{\epsilon_{\rm e}}{\epsilon_{\rm B}}\frac{1}{3-p}\left(\frac{\gamma_{\rm m}}{\gamma_{\rm c}^s}\right)^{p-2}\right]\left[p(1+Y)^{p-1}-\left(\frac{\gamma_{\rm m}}{\gamma_{\rm c}^s}\right)^{p-1}\right]^{-1}
\end{equation}

\section{Derivation of SSC with KN Suppression}\label{appKN}

Derivations involving KN effects are more approximate than the ones above, in part because the scattering cross-section is now dependent on the energy of the individual scatters. We derive here equations based on the bulk properties of the electron population, and assume a simplified version of the KN cross-section so that we can ignore effects due to individual photon scatterings. We also assume an optical depth $\tau_e \ll 1$ which is a reasonable assumption as a typical $n_0$ of order $1\ \rm cm^{-3}$ will yield $\tau_e\sim10^{-8}$, given the characteristic size associated with early afterglows of $10^{17}\ \rm cm$. This means multiple scatterings are sufficiently suppressed so that we can safely ignore them. Unlike in the Thomson scattering regime, we cannot assume that $Y$ is a simple function of $\langle \gamma_{\rm e}^2 \rangle$ because there is now a dependence on the incident photon energy. We follow a framework similar to the one by \citet{Nakar}, while including our more detailed description of $Y$ in the Thomson regime.

We define the synchrotron emissivity of a single electron, $P_{\nu}\left(\gamma^*\right)$, in Equation~\ref{Ykn:eq} as
\begin{equation}
\label{synchpow}
P_{\nu}\left(\gamma^{*}\right) \propto \begin{cases} 
       \delta\left(\nu-\nu\left(\gamma^{*}\right)\right)\nu(\gamma^{*})& \nu \gtrsim \nu\left(\gamma^{*}\right)\\
      \nu^{\frac{1}{3}}& \nu \ll \nu\left(\gamma^*\right)\\
   \end{cases}
\end{equation}
where the upper limit corresponds to the high-energy emission of the electron, and the lower limit corresponds to the low-energy synchrotron tail. We substitute Equation \ref{synchpow} into \ref{Ykn:eq} to obtain two equations for $Y$ depending on what portion of the photons can be Thomson scattered by $\gamma^*$ electrons:
\begin{equation}
\label{Ykn:low}
    Y\left(\gamma_{\rm e}\right) \ \propto \  \int_{0}^{\Tilde{\nu}\left(\gamma_{\rm e}\right)}d\nu' \nu'^{\frac{1}{3}}\int d\gamma_{\rm e}^*\frac{d n_0^{\prime}}{d\gamma_{\rm e}^*}
\end{equation}
and 
\begin{equation}
\label{Ykn:hi}
    Y\left(\gamma_{\rm e}\right) \ \propto \  \int_{0}^{\Tilde{\nu}\left(\gamma_{\rm e}\right)}d\nu'\int d\gamma_{\rm e}^*\delta\left(\nu'-\nu'\left(\gamma_{\rm e}^{*}\right)\right)\nu'(\gamma_{\rm e}^{*}) \frac{d n_0^{\prime}}{d\gamma_{\rm e}^*}
\end{equation}
\ref{Ykn:low} is a straightforward integration which results in $Y\ \propto\ \tilde{\nu}^{-\frac{4}{3}}$. For \ref{Ykn:hi}, we can exploit the fact that $\nu\ \propto\ \gamma_{\rm e}^2$, along with a property of the Dirac delta, to arrive at an equation for the high-energy scatterings:

\begin{equation}
\label{Ykn:hif}
    Y\left(\gamma_{\rm e}\right) \ \propto \  \int_{1}^{\Tilde{\gamma}\left(\gamma_{\rm e}\right)} d\gamma\gamma_{\rm e}^{2} \frac{d n_0^{\prime}}{d\gamma_{\rm e}}
\end{equation}

At this point, there are two ways to proceed. The simpler method is to determine the functional form of the major KN regimes, and then smoothly join them to the Thomson regime solution. This method requires only knowing the $\gamma_{\rm e}$ dependence of $Y$, and lets the simpler Thomson solution for $Y$ dictate the magnitude of $Y$. The second method is to compare \ref{Ykn:hif} to \ref{Y:gammal}, determine what constants are needed for \ref{Ykn:hif} to equal \ref{Y:gammal} in the Thomson regime, and then perform similar derivations to the ones found above. Doing the latter would give

\begin{equation}
\label{Ykn:hfull}
    Y\left(\gamma_{\rm e}\right) \ = \frac{4}{3}\sigma_{\rm T} n_0^{\prime} \Delta R^{\prime}  \frac{1}{n_0^{\prime}}\int_{1}^{\Tilde{\gamma}\left(\gamma_{\rm e}\right)} d\gamma_{\rm e}\gamma_{\rm e}^{2} \frac{d n_0^{\prime}}{d\gamma_{\rm e}}
\end{equation}
However, in our implementation we chose the former as it greatly simplified implementation in \texttt{boxfit} and allowed us to directly compare our results to \citet{Nakar}. Here we present the derivation used to determine the functional dependencies.

\subsection{Fast Cooling}

In the fast-cooling regime, we use Equation \ref{ne:fast} for the electron energy distribution and \ref{n0:fast} for $n_0$. Equation \ref{Ykn:low} yields one regime, while \ref{Ykn:hfull} yields two major regimes: $\tilde{\gamma}_{\rm e} > \gamma_{m}$ and $\tilde{\gamma}_{\rm e} < \gamma_{m}$.

\subsubsection{Weak KN Regime}

The weak KN regime, for which $\gamma_{\rm{c}} < \tilde{\gamma}_{\rm{e}} < \gamma_{\rm{m}}$, yields the main difference between the Thomson $Y$ and $Y\left(\nu_{\rm e},t\right)$ as the other regime $\left( \tilde{\gamma}_{\rm e} > \gamma_{\rm e} \right)$ very quickly returns to the Thomson $Y$. Substituting Equations \ref{ne:fast} and \ref{n0:fast} into \ref{Ykn:hfull}, and using the definitions of $\gamma_{\rm c}$ and $\gamma_{\rm m}$, leads to the following solution:
\begin{equation}
\label{Ykn:hifast}
    Y\left(\gamma_{\rm e}\right) \ = \frac{\epsilon_{\rm e} (p-2)}{\epsilon_{\rm B} (p-1)(1+Y_c)} \left[\left(\frac{\gamma_{\rm e}}{\hat{\gamma}_{\rm m}}\right)^{-\frac{1}{2}}-\frac{\gamma_{\rm c}}{\gamma_{\rm m}}\right]\left[1-\frac{1-p}{p}\frac{\gamma_{\rm c}}{\gamma_{\rm m}}\right]^{-1}
\end{equation}
Here we have introduced $Y_c=Y\left( \nu_{\rm c}\right)$. Taking the ultra fast-cooling limit results in $Y\ \propto \ \gamma^{-\frac{1}{2}}$, which we can connect to our Thomson solution at the boundary. One important thing to note is that although $Y$ goes to 0 at the boundary, this is an artifact of the approximation made for $P_{\nu}\left(\gamma_{\rm e}\right)$. To alleviate this, we use only the ultra fast-cooling approximation, so that we can smoothly connect this regime to Equation \ref{Ykn:strongf}.

\subsubsection{Transition to the Thomson Regime}

The derivation in this regime follows the same method as the one above. Since $\tilde{\gamma_{\rm e}} > \gamma_{\rm m}$, there are contributions from $\gamma^{-p-1}$ electrons:
\begin{equation}
\label{Ykn:tranfast}
    Y\left(\gamma_{\rm e}\right) \ = \frac{\epsilon_{\rm e} (p-2)}{\epsilon_{\rm B} (p-1)(1+Y_c)} \left[\frac{p-2}{p-1}-\frac{1}{p-2}\left(\frac{\gamma_{\rm e}}{\hat{\gamma}_{\rm m}}\right)^{\frac{p-2}{2}}-\frac{\gamma_{\rm c}}{\gamma_{\rm m}}\right]\left[1-\frac{1-p}{p}\frac{\gamma_{\rm c}}{\gamma_{\rm m}}\right]^{-1}
\end{equation}
In the limit $\hat{\gamma}_{\rm m} \gg \gamma_{\rm e}$, Equation \ref{Ykn:tranfast} reduces to \ref{Yt:fast}. Additionally, it agrees with \ref{Ykn:hifast} in the limit $\gamma_{\rm e} = \hat{\gamma}_{m}$.

\subsubsection{Strong KN Regime}

In this regime, for which $\tilde{\gamma}_{\rm{e}} < \gamma_{\rm{c}}$, $Y$ depends only on $\tilde{\nu}$, which can be rewritten in terms of $\gamma$ as
\begin{equation}
\label{Ykn:strongf}
    Y \ \propto \ \gamma_{\rm e}^{-4/3}
\end{equation}

To connect the three regimes, we approximate \ref{Ykn:tranfast} as the Thomson $Y$, then choose constants for Equations \ref{Ykn:hifast} and \ref{Ykn:strongf} such that they agree at the boundaries. These normalized equations are then used to solve for $Y$.

\subsection{Slow Cooling}

In the slow-cooling regime, we use Equation \ref{ne:slow} for the electron energy distribution and \ref{n0:slow} for $n_0$. As in the fast-cooling case, Equation \ref{Ykn:low} yields one regime, while \ref{Ykn:hfull} yields two major regimes: $\tilde{\gamma}_{\rm e} > \gamma_{c}$ and $\tilde{\gamma}_{\rm e} < \gamma_{c}$.

\subsubsection{Weak KN Regime}

Using the same methods as in the fast-cooling weak KN regime, but now for $\gamma_{\rm{m}} < \tilde{\gamma}_{\rm{e}} < \gamma_{\rm{c}}$, we can substitute Equations \ref{ne:slow} and \ref{n0:slow} into \ref{Ykn:hfull}, resulting in
\begin{equation}
\label{Ykn:weaks}
    Y=\frac{\epsilon_{\rm e}(p-2)}{\epsilon_{\rm B}(3-p)(1+Y_c)}\left[\left(\frac{\gamma_{\rm e}}{\hat{\gamma}_{\rm c}}\right)^{-\frac{1}{2}}\left(\frac{\gamma_{\rm e}}{\hat{\gamma}_{\rm m}}\right)^{\frac{p-2}{2}}-\frac{\gamma_{\rm m}}{\gamma_{\rm c}}\right]\left[1-\frac{1}{p}\left(\frac{\gamma_{\rm m}}{\gamma_{\rm c}}\right)^{p-1}\right]^{-1}
\end{equation}
This can be rewritten as 
\begin{equation}
\label{Ykn:weakalt}
    Y =\frac{\epsilon_{\rm e}(p-2)}{\epsilon_{\rm B}(3-p)(1+Y_c)}\left(\frac{\gamma_{\rm m}}{\gamma_{\rm c}}\right)^{p-2}\left[\left(\frac{\gamma_{\rm e}}{\hat{\gamma}_{\rm c}}\right)^{\frac{p-3}{2}}-\left(\frac{\gamma_{\rm m}}{\gamma_{\rm c}}\right)^{3-p}\right]\left[1-\frac{1}{p}\left(\frac{\gamma_{\rm m}}{\gamma_{\rm c}}\right)^{p-1}\right]^{-1}
\end{equation}
which brings it in line with the solution presented in \citet{Nakar}.

\subsubsection{Transition to the Thomson Regime}

In this regime, for which $\tilde{\gamma}_{\rm{e}} > \gamma_{\rm{c}}$, we gain contributions from $\gamma_{\rm e}^{-p-1}$ photons, which results in
\begin{equation}
\label{Ykn:trans}
    Y=\frac{\epsilon_{\rm e}}{\epsilon_{\rm B}(3-p)(1+Y_c)}\left[\left(\frac{\gamma_{\rm m}}{\gamma_{\rm c}}\right)^{p-2}-\frac{p-2}{3-p}\frac{\gamma_{\rm m}}{\gamma_{\rm c}}+(p-3)\left(\frac{\gamma_{\rm e}}{\hat{\gamma}_{\rm m}}\right)^{\frac{p-2}{2}}\right]\left[1-\frac{1}{p}\left(\frac{\gamma_{\rm m}}{\gamma_{\rm c}}\right)^{p-1}\right]^{-1}
\end{equation}
Here, we can also rewrite the solution:
\begin{equation}
\label{Ykn:translow}
    Y=\frac{\epsilon_{\rm e}}{\epsilon_{\rm B}(3-p)(1+Y_c)}\left(\frac{\gamma_{\rm m}}{\gamma_{\rm c}}\right)^{p-2}\left[1-\frac{p-2}{3-p}\left(\frac{\gamma_{\rm m}}{\gamma_{\rm c}}\right)^{3-p}+(p-3)\left(\frac{\gamma_{\rm e}}{\hat{\gamma}_{\rm c}}\right)^{\frac{p-2}{2}}\right]\left[1-\frac{1}{p}\left(\frac{\gamma_{\rm m}}{\gamma_{\rm c}}\right)^{p-1}\right]^{-1}
\end{equation}

\subsubsection{Strong KN Regime}

In this regime, with $\tilde{\gamma}_{\rm{e}} < \gamma_{\rm{m}}$, $Y$ depends only on $\tilde{\nu}$, which can be rewritten in terms of $\gamma$ as
\begin{equation}
\label{Ykn:strongs}
    Y \ \propto \ \gamma_{\rm e}^{-4/3}
\end{equation}

As in the fast-cooling case, to connect the three regimes, we approximate Equation \ref{Ykn:translow} as the Thomson $Y$, and then choose constants for Equations \ref{Ykn:weakalt} and \ref{Ykn:strongs} such that they agree at the boundaries. These normalized equations are used to solve for the $Y$ given above. Much like in the Thomson case, i.e. $Y_{\rm{T}}$ the transition between the fast- and slow-cooling regimes simplifies the electron population, resulting in 
\begin{equation}
\label{Ykn:tranfs}
Y\left(\gamma_{\rm e}\right) = Y_{\rm T} \begin{cases} 
      1 & \gamma_{\rm e} > \hat{\gamma}_*\\
      \left(\frac{\gamma_{\rm e}}{\hat{\gamma}_{*}}\right)^{-4/3}& \gamma_{\rm e} < \hat{\gamma}_*\\
   \end{cases}
\end{equation}
since there is no intermediate population between the two critical frequencies.


\label{lastpage}
\end{document}